\documentclass[amssymb,aps,prd,twocolumn,nofootinbib,superscriptaddress]{revtex4-2}
\usepackage{graphicx}
\usepackage{subcaption} 
\usepackage{mathtools}
\usepackage{slashed,bm}
\usepackage{url}
\usepackage[normalem]{ulem}
\usepackage{caption}
\usepackage{booktabs} 
\usepackage{enumitem}
\usepackage{mathrsfs}
\usepackage{float}
\usepackage{ragged2e}
\usepackage[colorlinks=true, linkcolor=blue, citecolor=red, urlcolor=magenta]{hyperref}
\usepackage{amsmath,amssymb,bm}
\usepackage{hyperref}
 \usepackage{pgfplotstable}
\usepackage{tikz,pgfplots}
\usetikzlibrary{pgfplots.groupplots} 
\pgfplotsset{compat=1.18}
\usepgfplotslibrary{fillbetween}
\usepgfplotslibrary{groupplots}
\usepackage{comment}
\usepackage{xcolor}
\definecolor{myblue}{RGB}{33,113,181}
\definecolor{myred}{RGB}{215,48,39}
\definecolor{mygreen}{RGB}{35,139,69}
\definecolor{mygold}{RGB}{217,95,2}
\definecolor{mypurple}{RGB}{117,107,177}
\definecolor{mygray}{RGB}{90,90,90}

\begin{document}

\title{Chiral Vortical Transport under Minimal-Length Deformations in Rotating Relativistic Matter}

\author{Sameer Ahmad Mir}
\email{sameerphst@gmail.com}
\affiliation{Canadian Quantum Research Center, 460 Doyle Ave 106, Kelowna, BC V1Y 0C2, Canada} 
\affiliation{Department of Computer Sciences, Asian School of Business, Noida, Uttar Pradesh, 201303, India}
\author{Behnam Pourhassan}
\affiliation{School of Physics, Damghan University, 3671641167 Damghan, Iran}
\affiliation{Center for Theoretical Physics, Khazar University, 41 Mehseti Street, Baku, AZ1096, Azerbaijan}
\author{Nawal H.Siddig}
\affiliation{Department of Mathematical Science, College of Science, Princess Nourah bint Abdulrahman University, P.O. Box 84428, Riyadh 11671, Saudi Arabia}
\author{Euaggelos E. Zotos}
\affiliation{Department of Physics, School of Science, Aristotle University of Thessaloniki, GR-541 24, Thessaloniki, Greece}
\affiliation{S.M. Nikolskii Mathematical Institute of the Peoples’ Friendship University of Russia (RUDN University), Moscow 117198, Russia}
\author{Laila Abdulaziz Al-Essa}
\affiliation{Department of Mathematical Science, College of Science, Princess Nourah bint Abdulrahman University, P.O. Box 84428, Riyadh 11671, Saudi Arabia}

\begin{abstract}
We study how a minimal length can affect the chiral vortical effect, in which rotation produces a current in matter made of chiral fermions. We focus on an isotropic modification of the momentum-space density-of-state (DoS) and determine the vortical conductivity from the static, long-wavelength correlation between the current and momentum density. The analysis is carried out first for a single right-handed Weyl fermion and is then extended to general massless chiral fermion systems with mutually commuting conserved charges. We find that, for a broad class of analytic isotropic DoS modifications, the familiar leading contributions from chemical potential and temperature remain unchanged at first order within the DoS sector. Minimal-length effects instead appear through higher-order thermal and chemical-potential contributions. Some of these corrections contain the same chiral charge combinations that appear in gauge and mixed gauge-gravitational anomalies, while another contains an independent higher-order charge combination. Their coefficients depend on the form of the DoS modification and are therefore not fixed by the usual four-dimensional anomaly structure. For the Kempf-Mangano-Mann deformation, the exact result shows that a positive minimal-length parameter always reduces the DoS contribution to the vortical conductivity. We also find that a Planck-suppressed correction is extremely small under standard heavy-ion chemical freeze-out conditions. These results show that isotropic minimal-length modifications can change higher-order contributions to chiral vortical transport while leaving the familiar leading anomaly-related response unchanged within the DoS sector considered here.
\end{abstract}

\maketitle

\section{Introduction}
 
Ultra-relativistic heavy-ion collisions create short-lived droplets of quantum chromodynamics (QCD) matter under extreme conditions of temperature and density, while noncentral collisions additionally endow the system with a large orbital angular momentum
\cite{ALICE:2010khr,ALICE:2010suc,Becattini:2007sr,Becattini:2015ska,
STAR:2005gfr,PHENIX:2004vcz}. During the subsequent evolution, part of
this initial orbital angular momentum is transferred to the expanding
medium, giving rise to fluid vorticity and thermal vorticity
\cite{Becattini:2007sr,Becattini:2015ska}.
The observation of global \(\Lambda\) and \(\bar{\Lambda}\) polarization has provided direct experimental evidence that the matter produced in such collisions is a strongly vortical relativistic fluid \cite{STAR:2017ckg,STAR:2018gyt,STAR:2023nvo,STAR:2021beb,HADES:2022enx,ALICE:2019onw}. This experimental access to rotation has sharpened interest in anomalous transport phenomena that couple directly to vorticity \cite{Vilenkin:1979ui,Son:2009tf,Landsteiner:2011cp,Kharzeev:2015znc}. At the same time, it is important to keep conceptually distinct the microscopic transport coefficients of a chiral medium \cite{Landsteiner:2011cp,Landsteiner:2012kd} and experimentally measured spin observables \cite{Becattini:2013fla,Becattini:2016gvu,Becattini:2022zvf}: global hyperon polarization establishes the presence of a vortical environment, but it is not by itself a direct measurement of the chiral vortical conductivity. Among anomaly-induced transport phenomena, the chiral vortical effect (CVE) provides a particularly important connection between quantum anomalies and the response of rotating relativistic matter \cite{Son:2009tf,Kharzeev:2010gr,Landsteiner:2011iq,Liu:2018xip}. In a chiral medium, rotation induces vector or axial currents along the vorticity direction, and in equilibrium the leading coefficients of this response are tied to gauge and mixed gauge-gravitational anomalies rather than to ordinary dissipative transport physics \cite{Vilenkin:1979ui,Son:2009tf,Landsteiner:2011cp,Landsteiner:2016led,Pourhassan:2025txr}. This is what makes the CVE conceptually powerful: it connects a macroscopic transport response of the medium to microscopic anomaly data \cite{Son:2009tf,Landsteiner:2011iq,Jensen:2012kj}. Consequently, the leading anomaly-induced vortical structures are strongly constrained by the underlying symmetry and anomaly content of the theory \cite{Banerjee:2012iz,Landsteiner:2011iq}. So long as the relevant symmetries are preserved and no new chiral degrees of freedom are introduced, the universal leading vortical response should remain fixed, even though its phenomenological manifestation in heavy-ion collisions must still contend with finite lifetime, hadronization, feed-down, and freeze-out effects \cite{Kharzeev:2015znc,Wang:2016mkm,Wang:2023xhn}.

Against this background, we investigate how the anomaly-controlled
vortical response is modified by an isotropic minimal-length
deformation of the momentum-space DoS
~\cite{Kempf:1994su,Kempf:1996nk,Chang:2001bm}.
A natural framework for addressing this question is the Generalized
Uncertainty Principle (GUP), in which the canonical
position-momentum algebra receives momentum-dependent corrections
\cite{Kempf:1994su,Chang:2001bm}.
At the semiclassical level relevant here, the minimal-length
deformation changes the momentum-space DoS and hence
the weighting of equilibrium momentum integrals
\cite{Chang:2001bm,Fityo:2008zz}.
The CVE therefore provides a controlled framework for examining
whether this ultraviolet modification affects only regular thermal
contributions or also the leading anomaly-controlled vortical
structures \cite{Son:2009tf,Landsteiner:2011cp,Jensen:2012kj,Banerjee:2012iz}. 
There are several reasons why this question is theoretically well motivated. First, anomaly-induced transport provides a stringent consistency test for ultraviolet modifications of the theory. The leading vortical coefficients are constrained by the gauge and mixed gauge-gravitational anomaly data \cite{Son:2009tf,Landsteiner:2011cp,Jensen:2012kj,Banerjee:2012iz}. Accordingly, a deformation that leaves the chiral field content and the corresponding anomaly coefficients unchanged should reproduce the same anomaly-controlled leading structures, while any additional contributions must arise from sectors not fixed by the anomaly
polynomial. 
Second, the CVE admits a direct Kubo formulation in terms of the mixed current-momentum-density correlator, providing a microscopic definition of the vortical conductivity
\cite{Amado:2011zx,Landsteiner:2011cp,Landsteiner:2012kd}. This formulation is particularly useful here because it allows the anomaly-controlled contribution to be distinguished from changes in the regular thermal momentum integrals induced by the deformed DoS. 
Third, although Planck-suppressed minimal-length corrections are negligible at Relativistic Heavy-Ion Collider (RHIC) and Large Hadron Collider (LHC) energies in the present analysis, their analytic form remains informative. It identifies the higher thermal structures induced by the deformed DoS and provides a systematic basis for comparing quantum-gravity motivated modifications
of relativistic thermodynamics \cite{Chang:2001bm,Vakili:2012qt,AbouSalem:2015wha}.
Recent work has also examined minimal-length effects directly in
effective descriptions of chiral quark matter, including the
\(2+1\)-flavor Nambu-Jona-Lasinio framework
\cite{Mir:JPhysG}.

In this work, we isolate the isotropic DoS sector of a
minimal-length deformation while retaining the massless Weyl dispersion
relation, current operator, and Belinfante energy-momentum tensor in
their undeformed free-theory forms
\cite{Kempf:1994su,Chang:2001bm,Fityo:2008zz}. We work in natural
units. Throughout the paper, the scalar $\beta$, also denoted as the
GUP deformation parameter, characterizes the minimal-length
modification and has dimensions of inverse energy squared
\cite{Kempf:1994su,Chang:2001bm}. The unindexed symbol $\beta$ is used
only for this deformation parameter and never for inverse temperature.
The equilibrium temperature is denoted by $T$, while $\mu$ denotes the
chemical potential associated with the $U(1)$ charge of the Weyl
fermion. In the covariant description of relativistic equilibrium, the
inverse-temperature four-vector is denoted by $\beta^\mu(x)$, or by
$b^\mu$ in global equilibrium. The Lorentz index therefore
unambiguously distinguishes the inverse-temperature four-vector from
the scalar minimal-length parameter. The present analysis isolates the
DoS contribution, possible modifications of propagators,
source vertices, contact terms, and Ward identities in a complete
source-covariant microscopic formulation are considered separately in
Sec.~\ref{sec:consistency_anomalies}. 
For the unit-charge right-handed Weyl fermion considered first,
$J^\mu$ denotes the $U(1)$ current and $T^{\mu\nu}$ denotes the
symmetric Belinfante energy-momentum tensor. Their components $J^i$
and $T^{0j}$ represent the spatial current and momentum density,
respectively. At first order in fluid vorticity, the spatial current is
proportional to the vorticity, with the proportionality coefficient
defined as the chiral vortical conductivity $\sigma_V$. This
conductivity is obtained from the parity-odd part of the retarded mixed
current-momentum density correlator in the static,
long-wavelength limit
\cite{Amado:2011zx,Landsteiner:2011cp,Landsteiner:2012kd}. It is
therefore an equilibrium transport coefficient and should not be
identified directly with a hadronic spin-polarization observable. 
The Kempf-Mangano-Mann (KMM) form provides the explicit DoS modification used
for the exact finite-deformation analysis
\cite{Kempf:1994su,Chang:2001bm,Fityo:2008zz}. The symmetry argument
underlying the parity-odd tensor reduction is more general, however,
because it requires only rotational invariance of the momentum-space
modification. We therefore extend the first-order analysis to a broad
analytic isotropic class in which the DoS factor depends
only on the dimensionless combination formed from the deformation
parameter and the squared magnitude of the three-momentum. Near the
undeformed limit, the first correction is proportional to this
combination, with a model-dependent coefficient determined by the
slope of the DoS function at the origin. For the KMM
choice, this coefficient is \(c_1=-3\) \cite{Kempf:1994su,Chang:2001bm,Fityo:2008zz}. Because the
DoS factor is a rotational scalar, it modifies only the
radial thermal moments and cannot introduce an additional parity-odd
tensor structure. The antisymmetric structure of the response remains
that of the mixed $J^iT^{0j}$ correlator, as demonstrated explicitly in
Sec.~\ref{sec3} and App.~\ref{App:a}
\cite{Amado:2011zx,Landsteiner:2011cp,Landsteiner:2012kd}. 
Within this analytic isotropic DoS class, the familiar
leading contributions to the vortical conductivity that are quadratic
in chemical potential and temperature retain their undeformed
coefficients through first order in the minimal-length parameter. This
is consistent with the conventional anomaly-related structure of the 
CVE, in which the chemical-potential contribution
is associated with the gauge anomaly and the thermal contribution with
the mixed gauge-gravitational anomaly
\cite{Landsteiner:2011cp,Landsteiner:2011iq,Jensen:2012kj,
Golkar:2012kb,Golkar:2015oxw}. Minimal-length dependence first appears
at the next thermodynamic order. The corresponding corrections contain
a contribution quartic in the chemical potential, a mixed term
quadratic in both chemical potential and temperature, and a purely
thermal term quartic in temperature. Their overall fourth-order form
is fixed by dimensional consistency, but dimensional arguments alone
cannot determine their numerical coefficients, relative signs, or
chiral charge dependence. These properties follow from the explicit
parity-odd Kubo calculation
\cite{Neiman:2010zi,Loganayagam:2012pz,Landsteiner:2012kd,Yee:2014dxa}.

For a general chiral multiplet, the mixed
chemical-potential thermal correction carries the same cubic chiral
charge trace that appears in the gauge-anomaly structure, while the
purely thermal correction carries the linear chiral trace associated
with the mixed gauge-gravitational anomaly. The quartic
chemical-potential correction instead contains an independent
fifth-order chiral charge trace
\cite{Neiman:2010zi,Loganayagam:2012pz,Stone:2018zel}. The appearance
of these traces does not imply that the higher-order coefficients are
fixed by the anomaly polynomial. Their numerical values depend on the
chosen DoS deformation and therefore represent
model-dependent higher thermal contributions. This distinction is
consistent with equilibrium formulations in which the leading
hydrostatic anomalous response is constrained by the anomaly content
of the underlying chiral theory
\cite{Banerjee:2012iz,Jensen:2012kj,Jensen:2013kka}. The exact
monotonic reduction of the DoS contribution for positive
deformation is a property of the KMM form, whereas the first-order
classification in terms of chiral charge traces applies to the broader
analytic isotropic class considered here. 
For phenomenological orientation, Sec.~\ref{sec:phenomenology} evaluates the
equilibrium result along standard chemical freeze-out relations
between \(\sqrt{s_{NN}}\), \(\mu_B\), and \(T\)
\cite{Cleymans:2005xv,Andronic:2017pug}.
Global-polarization measurements are used only to establish the
presence and beam-energy dependence of a vortical medium, not as a
direct determination of \(\sigma_V\)
\cite{STAR:2017ckg,Becattini:2016gvu,Karpenko:2016jyx}.
For a Planck-suppressed deformation, the resulting minimal-length
contribution remains negligible under heavy-ion conditions.
The precise DoS framework used in the analysis is
defined in Sec.~\ref{sec2}. Its relation to a more general
source-covariant microscopic theory, including possible propagator,
vertex, contact, and Ward-identity contributions, is examined in
Sec.~\ref{sec:consistency_anomalies}.

The remainder of the paper is organized as follows. Sec.~\ref{sec2} introduces the semiclassical GUP framework used throughout the analysis, including the isotropic modification of the momentum-space DoS and the equilibrium description of rotating relativistic matter. Sec.~\ref{sec3} derives the chiral vortical conductivity for a massless right-handed Weyl fermion from the static, long-wavelength mixed (\(J^iT^{0j}\)) Kubo correlator and identifies the leading minimal-length contribution to the thermal moments. Sec.~\ref{sec:consistency_anomalies} examines the relation between these contributions and anomaly physics, extends the result to general massless chiral multiplets, and makes explicit the roles of the gauge-anomaly trace (\(d_{ABC}\)), the mixed gauge-gravitational trace (\(b_A\)), and higher chiral charge traces. The same section also clarifies how the DoS contribution would enter a more complete source-covariant formulation containing additional propagator, vertex, contact, and Ward-identity terms. Sec.~\ref{sec:phenomenology} distinguishes vector and axial vortical currents from momentum-dependent spin observables and uses chemical freeze-out relations to assess the size of the minimal-length contribution under heavy-ion conditions. Sec.~\ref{sec6} summarizes the principal results and their domain of validity. App.~\ref{App:a} provides the detailed one-loop evaluation of the mixed correlator, including the parity-odd projection, Matsubara summation, vacuum subtraction, and static long-wavelength limit.

\section{GUP-Deformed Phase Space and Equilibrium Distribution}
\label{sec2}
\noindent
In this section, we introduce the semiclassical GUP formulation used throughout the analysis. We isolate the isotropic minimal-length contribution associated with the momentum-space DoS that appears in the equilibrium integrals relevant to the vortical Kubo relation \cite{Kempf:1994su,Chang:2001bm,Fityo:2008zz}.
The underlying chiral fermion content, dispersion relation, current operator, and Belinfante energy-momentum tensor retain their undeformed free-theory forms. This choice separates the modification of the thermal momentum moments from possible changes of propagators, source vertices, contact terms, and Ward identities that can arise in more complete GUP field-theory formulations \cite{Husain:2012im}. 
We adopt the isotropic quadratic KMM algebra, for which the modified position-momentum relation leads to a momentum-dependent DoS \cite{Kempf:1994su,Chang:2001bm}.
Within the present formulation, the minimal-length dependence therefore
appears only through the radial thermal moments. Whether the standard
anomaly-controlled contributions to the vortical conductivity retain
their undeformed coefficients is examined explicitly in
Sec.~\ref{sec3} through the mixed current-momentum-density Kubo
correlator
\cite{Amado:2011zx,Landsteiner:2011cp,Landsteiner:2012kd}. 
For the rotationally invariant sector considered below, we denote the magnitude of the three-momentum by $P\equiv|\mathbf{P}|$. We take as the starting point the isotropic quadratic branch of the KMM algebra, whose semiclassical limit may be written as
\begin{equation}
\{X_i,P_j\}
=
\delta_{ij}\left(1+\beta P^2\right)
+\mathcal O(\beta^2),
\,\,
\{P_i,P_j\}=0.
\label{eq:PB}
\end{equation}
Here and throughout the paper,
$\beta\equiv\beta_{\rm GUP}>0$ is the scalar KMM
minimal-length parameter and has dimensions of inverse energy squared,
$[\beta]=E^{-2}$ \cite{Kempf:1994su,Kempf:1996nk,Chang:2001bm}.
It is never used to denote the inverse temperature $1/T$.
In relativistic-equilibrium expressions, we reserve
$\beta^\mu(x)=u^\mu(x)/T(x)$ for the inverse-temperature four-vector
and $b^\mu$ for its constant global-equilibrium value. The second KMM
deformation parameter is denoted by $\beta'$ and is set to zero in the
isotropic branch considered here
\cite{Kempf:1994su,Chang:2001bm}. 
For the rotationally invariant
momentum integrals considered below, the relevant semiclassical
information is contained in the Jacobian associated with the modified
DoS. The detailed operator representation and the
higher-order position-position brackets do not contribute to the
sector considered here
\cite{Kempf:1994su,Chang:2001bm,Fityo:2008zz}. Following the
semiclassical statistical formulation of KMM-type algebras, we use the
Liouville-invariant volume element \cite{Chang:2001bm,Fityo:2008zz},
\begin{equation}
d\Gamma_{\rm inv}
=
d^3X\,
\frac{d^3P}{\bigl(1+\beta P^2\bigr)^3}
\equiv
d^3X\,d^3P\,W_\beta(P),
\label{eq:LiouvilleMeasure}
\end{equation}
where $W_\beta(P)\equiv\frac{1}{\bigl(1+\beta P^2\bigr)^3}$. 
The cubic power reflects the three spatial momentum directions of the
isotropic KMM branch \cite{Chang:2001bm,Fityo:2008zz}. 
{The factor \(W_\beta(P)\) is applied after the free Weyl theory and
its source operators have been specified. Accordingly, the
propagator poles, current insertion, Belinfante stress-tensor
insertion, chirality, and charge assignment retain their undeformed
forms. Possible source-dependent propagator, vertex, contact, and
Ward-identity contributions beyond this DoS sector are
organized explicitly in Sec.~\ref{sec:consistency_anomalies}.
}

For weak deformation, Eq.~\eqref{eq:LiouvilleMeasure} takes the form
\begin{multline}
W_\beta(P)
=
1-3\beta P^2+\mathcal O(\beta^2),
\quad\Longrightarrow\quad \\
d\Gamma_{\rm inv}
=
d^3X\,d^3P\Bigl(1-3\beta P^2+\mathcal O(\beta^2)\Bigr).
\label{eq:MeasureExpansion}
\end{multline}

With the microscopic assumptions fixed above, we next specify the equilibrium state of the rotating relativistic system. 
Global thermodynamic equilibrium in relativistic quantum field theory is described by a covariant density operator involving the conserved four-momentum, Lorentz generators, and internal conserved charges \cite{Becattini:2016stj,Buzzegoli:2018wpy,Palermo:2021hlf}. In flat spacetime, this density operator may be written as \cite{Becattini:2016stj,Buzzegoli:2018wpy}
\begin{equation}
\hat\rho
=
\frac{1}{Z}
\exp\!\left[
-\,b_\mu \hat P^\mu
+\frac{1}{2}\varpi_{\mu\nu}\hat J^{\mu\nu}
+\zeta\,\hat Q
\right],
\label{eq:rho_general}
\end{equation}
{where $b^\mu$ is the constant inverse-temperature four-vector associated with global equilibrium, $\varpi_{\mu\nu}$ is the constant thermal-vorticity tensor, and $\zeta\equiv\mu/T$ is the dimensionless chemical-potential variable. The local inverse-temperature four-vector is written as $\beta^\mu(x)=u^\mu(x)/T(x)$, where $u^\mu(x)$ is the fluid four-velocity. The Lorentz-indexed quantity $\beta^\mu(x)$ is therefore distinct from the scalar minimal-length parameter $\beta\equiv\beta_{\rm GUP}$ introduced in
Eqs.~\eqref{eq:PB}-\eqref{eq:MeasureExpansion}.} 
For rigid rotation about the \(z\)-axis, \(\bm\Omega=\Omega\,\hat{\bm z}\), and within the local region \( |\Omega|\,r \ll 1 \) in which rigid rotation is well defined, one may choose the comoving frame \(b^\mu=(1/T,0,0,0)\) and \(\varpi_{12}=-\varpi_{21}=\Omega/T\), with all other independent components vanishing. Eq.~\eqref{eq:rho_general} then reduces to
\begin{equation}
\hat\rho
\propto
\exp\!\left[
-\frac{1}{T}\bigl(\hat H-\mu\hat N-\Omega \hat J_z\bigr)
\right],
\label{eq:rho_rot}
\end{equation}
where \(\hat J_z\) is the conserved total angular momentum along the rotation axis. For spin-\(\tfrac{1}{2}\) fermions, the local-equilibrium distribution matrix can be written compactly as \cite{Becattini:2013fla}. 
{Here we define
\begin{equation}
\xi_p(x)\equiv \beta_\mu(x)p^\mu-\zeta(x),
\qquad
f_0(\xi_p)\equiv \frac{1}{e^{\xi_p}+1},
\end{equation}
the spin-dependent local-equilibrium distribution required at the
order considered here is
\begin{multline}
f_{\rm eq}(x,p)
=
f_0(\xi_p)\,\mathbf{1}
+
\frac{1}{2}
f_0(\xi_p)\bigl[1-f_0(\xi_p)\bigr]\times
\\
\varpi_{\mu\nu}(x)\Sigma^{\mu\nu}
+
\mathcal O(\varpi^2),
\quad
\Sigma^{\mu\nu}
=
\frac{i}{4}[\gamma^\mu,\gamma^\nu].
\label{eq:linear_spin_distribution}
\end{multline}
Eq.~\eqref{eq:linear_spin_distribution} is used only to first order in thermal vorticity.  No statement about the distribution at second or higher order in rotational gradients is inferred from this form.  Exact equilibrium in pure rotation requires the full rotating thermal field theory, as originally given by Vilenkin~\cite{Vilenkin:1980zv} and, in modern covariant form, through exact global-equilibrium Wigner functions and spin observables for the Dirac field~\cite{Palermo:2021hlf}.
}
For a stationary rigidly rotating ensemble, this implies that the occupation of a one-particle state of energy \(E\) and angular-momentum eigenvalue \(j_z\) is biased according to
\begin{multline}
f_\Omega(E,j_z)
=
\frac{1}{e^{(E-\mu-\Omega j_z)/T}+1}
=\\
f_0(E,\mu,T)
+
\frac{\Omega}{T}\,j_z\,f_0(1-f_0)
+\mathcal O(\Omega^2),
\label{eq:fOmega_final}
\end{multline}
with
\begin{equation}
f_0(E,\mu,T)
=
\frac{1}{e^{(E-\mu)/T}+1}.
\label{eq:f0_def}
\end{equation}
Eq.~\eqref{eq:fOmega_final} gives the linearized equilibrium
relation associated with the coupling of rotation to the total angular
momentum in the grand-canonical ensemble. In the present analysis,
however, the vortical conductivity is obtained from the mixed
current-momentum-density Kubo correlator. Vorticity is therefore
implemented as a weak external rotational source, while the thermal
loop integrals are evaluated about the nonrotating equilibrium
ensemble. Equations~\eqref{eq:rho_rot}-\eqref{eq:fOmega_final} thus
establish the thermodynamic meaning of the rotating state and the
angular-momentum coupling; they do not substitute for the Kubo
derivation of the parity-odd transport coefficient
\cite{Amado:2011zx,Landsteiner:2011cp,Landsteiner:2012kd}. 

Thermal expectation values are subsequently evaluated using the
GUP-modified DoS. After factoring out the spatial volume,
the equilibrium expectation value of a single-particle observable
\(\mathcal O(\bm P)\) that is scalar in spin space takes the form
\begin{equation}
\langle \mathcal O\rangle_{\beta}
=
\int\frac{d^3P}{(2\pi)^3}\,
W_\beta(P)\,
\mathcal O(\bm P)\,
f_0(E,\mu,T)
\;+\;(\text{antiparticles}),
\label{eq:Oavg_GUP}
\end{equation}
where the antiparticle contribution is obtained by \(\mu\to -\mu\). Expanding to first order in the deformation gives
\begin{multline}
\langle \mathcal O\rangle_{\beta}
=
\int\frac{d^3P}{(2\pi)^3}\,
\Bigl(1-3\beta P^2\Bigr)\,
\mathcal O(\bm P)\,
f_0(E,\mu,T)
\\(\text{antiparticles})
\;+\;\mathcal O(\beta^2).
\label{eq:Oavg_expand}
\end{multline}
Within the semiclassical DoS framework adopted here, the
GUP dependence of equilibrium observables is carried solely by the
rotationally invariant modification of the momentum space density of
states. For the Kubo analysis below, it is useful to introduce the
corresponding deformed Fermi-Dirac moments. For the massless Weyl
fermion considered in the next section, the on-shell dispersion relation
is \(E=P\), and we define
\begin{multline}
I_n(\mu,T,\beta)
\equiv
\int_0^\infty \frac{dP}{2\pi^2}\,
P^n\,W_\beta(P)\,n_F(P-\mu),
\\
n_F(x)\equiv \frac{1}{e^{x/T}+1},
\label{eq:In_def}
\end{multline}
together with the antiparticle moments
\begin{equation}
\bar I_n(\mu,T,\beta)\equiv I_n(-\mu,T,\beta)
=
\int_0^\infty \frac{dP}{2\pi^2}\,
P^n\,W_\beta(P)\,n_F(P+\mu).
\label{eq:Ibar_def}
\end{equation}
At \(\beta=0\), one recovers the standard polylogarithmic form
\begin{equation}
I_n(\mu,T;0)
=
-\frac{\Gamma(n+1)}{2\pi^2}\,
T^{n+1}\,
\operatorname{Li}_{n+1}\!\bigl(-e^{\mu/T}\bigr),
\label{eq:polylog}
\end{equation}
while the weak-deformation expansion follows immediately from Eq.~\eqref{eq:MeasureExpansion}
\begin{equation}
I_n(\mu,T,\beta)
=
I_n(\mu,T,0)
-
3\beta\,I_{n+2}(\mu,T,0)
+
\mathcal O(\beta^2).
\label{eq:In_expand}
\end{equation}
These Fermi-Dirac moments provide the radial thermal integrals that appear in the one-loop Kubo correlator discussed in the next section. Since \(W_\beta(P)\) is a scalar function of \(P^2\), the minimal-length modification preserves rotational invariance and introduces no independent parity-odd tensor structure \cite{Kempf:1994su,Chang:2001bm,Fityo:2008zz}. Within the semiclassical DoS framework considered here, the GUP dependence of the vortical conductivity is therefore carried entirely by the modified thermal moments. 
The antisymmetric tensor structure associated with the chiral vortical conductivity instead originates from the parity-odd sector of the mixed \(J^iT^{0j}\) correlator, as required by the corresponding Kubo relation \cite{Amado:2011zx,Landsteiner:2011cp,Landsteiner:2012kd}.


\section{Vortical Current in Rotating Equilibrium}
\label{sec3}
\noindent
This section derives the chiral vortical conductivity \(\sigma_V\)
within the semiclassical DoS framework introduced in
Sec.~\ref{sec2}. For a single right-handed Weyl fermion of unit charge,
\(\sigma_V\) is defined by the first-order constitutive relation
\(J^i=\sigma_V\,\omega^i,\)
and is obtained from the parity-odd part of the mixed
current-momentum density correlator in the static, long-wavelength Kubo limit
\cite{Amado:2011zx,Landsteiner:2011cp,Landsteiner:2012kd}. 
The conductivity defined here is distinct from hadronic spin
polarization; the corresponding observable-level distinction is
discussed in Sec.~\ref{sec:phenomenology}. 
Accordingly, the present section is restricted to the equilibrium
Kubo derivation of \(\sigma_V\). Its beam-energy dependence and
heavy-ion implications are considered separately in
Sec.~\ref{sec:phenomenology}. 
A stationary slow rotation can be represented as a weak gravitomagnetic perturbation of the metric,
\begin{equation}
g_{\mu\nu}=\eta_{\mu\nu}+h_{\mu\nu},
\,\,
h_{0i}(\bm{x})=-(\bm{\Omega}\times \bm{x})_i,
\,\,
|h_{\mu\nu}|\ll 1,
\label{eq:gravimag}
\end{equation}
which couples to the momentum density \(T^{0i}\). Linear response then gives
\begin{multline}
\delta\langle J^i(x)\rangle
=
\frac{1}{2}\int d^4y\,
G_{JT}^{R\,ij}(x-y)\,h_{0j}(y),
\\
G_{JT}^{R\,ij}(x-y)
=
-i\,\Theta(x^0-y^0)\,
\langle [J^i(x),T^{0j}(y)]\rangle .
\label{eq:linresp}
\end{multline}
For a spatially uniform vorticity field, the induced chiral current takes the form
\begin{equation}
J^i=\sigma_V\,\omega^i,
\quad
\omega^i
\equiv
\frac{1}{2}\epsilon^{ijk}\partial_j u_k
\simeq
\frac{1}{2}\epsilon^{ijk}\partial_j v_k,
\label{eq:vorticity_def}
\end{equation}
so that for rigid rotation \(\bm v=\bm\Omega\times\bm x\) one has \(\nabla\times\bm v=2\bm\Omega\), hence \(\bm\omega=\bm\Omega\).

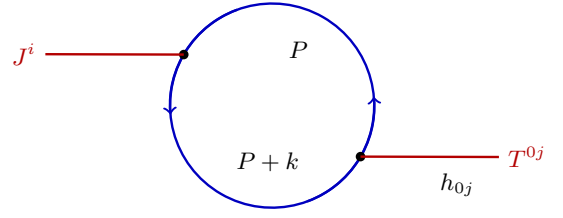
\begin{figure}[htb]
\centering
\begin{tikzpicture}[scale=1.0,line width=0.9pt]
  \def\r{1.35}

  \coordinate (A) at (150:\r);
  \coordinate (B) at (-30:\r);

  \draw[blue!75!black] (A) arc[start angle=150,end angle=330,radius=\r];
  \draw[blue!75!black] (B) arc[start angle=-30,end angle=150,radius=\r];

  \draw[->,blue!75!black] (110:\r) arc[start angle=110,end angle=185,radius=\r];
  \draw[->,blue!75!black] (290:\r) arc[start angle=290,end angle=365,radius=\r];

  \fill[black] (A) circle (1.8pt);
  \fill[black] (B) circle (1.8pt);

  \draw[red!70!black] (-3.0,0.68) -- (A);
  \draw[red!70!black] (B) -- (3.0,-0.68);

  \node[left,red!70!black] at (-3.0,0.68) {$J^i$};
  \node[right,red!70!black] at (3.0,-0.68) {$T^{0j}$};
  \node[above left] at (40:0.78) {$P$};
  \node[below right] at (220:0.78) {$P+k$};

  \node[right,font=\small] at (2.1,-1.05) {$h_{0j}$};

  \node[align=center,font=\small,teal!70!black] at (0,-2.35)
  {$\displaystyle 
  \int \frac{d^3P}{(2\pi)^3}
  \;\longrightarrow\;
  \int \frac{d^3P}{(2\pi)^3}\,W_\beta(P),
  \qquad
  W_\beta(P)=\frac{1}{(1+\beta P^2)^3}$};
\end{tikzpicture}
\caption{\justifying Schematic one-loop two-point function underlying the Kubo extraction of the chiral vortical conductivity. The \(U(1)\) current vertex \(J^i\) is correlated with the momentum-density vertex \(T^{0j}\), with the weak gravitomagnetic source \(h_{0j}\) supplying the vortical perturbation. The conductivity is isolated from the parity-odd part of this correlator in the static, long-wavelength limit \cite{Amado:2011zx,Landsteiner:2011iq,Landsteiner:2012kd,Yee:2014dxa}. Within the semiclassical GUP framework used here, the minimal-length dependence appears solely through the scalar DoS factor \(W_\beta(P)\) multiplying the spatial momentum integral, while the current and Belinfante stress-tensor vertices retain their undeformed free-theory forms \cite{Chang:2001bm,Fityo:2008zz,Vakili:2012qt}.
}
\label{fig:kubo_bubble}
\end{figure} 
Fig.~\ref{fig:kubo_bubble} summarizes the field-theoretic structure of the derivation. The chiral vortical conductivity is determined by the parity-odd component of the mixed current-momentum density correlator, whereas the minimal-length dependence appears solely through the
isotropic DoS factor \(W_\beta(P)\) multiplying the
spatial momentum integral \cite{Amado:2011zx,Landsteiner:2012kd,Landsteiner:2013aba,Yee:2014dxa}.
To isolate the vortical conductivity, we choose the external spatial
momentum along the \(z\)-direction,
\(\bm{k}=k\,\hat{\bm z}\). This choice projects the antisymmetric
transverse components of the retarded correlator onto the
Levi-Civita tensor structure characteristic of anomalous vortical
transport
\cite{Amado:2011zx,Landsteiner:2013aba,Yee:2014dxa}.
The conductivity is then extracted from the static,
long-wavelength limit of the parity-odd retarded correlator through
\begin{equation}
\sigma_V
=
\lim_{k\to 0}
\frac{i}{2k}\,
\epsilon^{ij3}\,
G^R_{J^iT^{0j}}(\omega=0,k\hat{\bm z}).
\label{eq:Kubo_sigmaV}
\end{equation}
{ The antisymmetric projector in Eq.~\eqref{eq:Kubo_sigmaV} isolates the parity-odd
vortical channel, but this tensor property is not by itself a proof
of nondissipation.  The nondissipative character follows from the
hydrostatic nature of the response: the vortical term survives in a
stationary equilibrium state and can be obtained from the equilibrium
generating functional~\cite{Landsteiner:2012kd,Banerjee:2012iz}.  When
microscopic time-reversal symmetry applies, both the spatial current
$J^i$ and the vorticity $\omega^i$ are odd under time reversal, so the
static response coefficient multiplying $J^i=\sigma_V\omega^i$ is
time-reversal even.  This should be contrasted with a dissipative
coefficient such as the shear viscosity, which is determined by the
absorptive low-frequency part of a retarded correlator,
\begin{equation}
\eta
=
-\lim_{\omega\rightarrow0}
\frac{1}{\omega}
\,{\rm Im}\,
G^R_{T^{xy}T^{xy}}(\omega,\mathbf 0).
\end{equation}
Thus the antisymmetric projector identifies the vortical tensor
structure, whereas its nondissipative character follows from its equilibrium and discrete-symmetry properties.} 
For the one-loop derivation, we embed the right-handed Weyl fermion in a
four-component Dirac representation and impose chirality through the
projector \(P_R=(1+\gamma^5)/2\), which keeps the parity-odd Dirac
structure explicit throughout the analysis
\cite{Landsteiner:2013aba,Yee:2014dxa,Hongo:2019rbd}.
With the external momentum chosen as
\(\bm{k}=k\hat{\bm z}\), the Euclidean mixed two-point function is written
in the standard finite-temperature form. The Belinfante
\(T^{0j}\) vertex then separates naturally into the terms proportional
to \(\gamma^0 p^j\) and \(\gamma^j i\tilde{\omega}_n\), whose individual
contributions are displayed in App.~\ref{App:a}
\cite{Landsteiner:2012kd,Landsteiner:2013aba,Hongo:2019rbd}.
After carrying out the Dirac trace, angular average, fermionic Matsubara
sum, and subtraction of the temperature- and chemical-potential independent
vacuum contribution, the static parity-odd sector of the mixed correlator
reduces to
\begin{equation}
G^{ij}_{E,\mathrm{odd}}(0,k\hat z)= i \epsilon^{ij3} k\, I_\beta(\mu,T)+O(k^2),
\end{equation}
where
\begin{multline}
I_\beta(\mu,T)=\frac{1}{2\pi^2}\int_0^\infty dP\, P\, W_\beta(P)\times\\\left[n_F(P-\mu)+n_F(P+\mu)\right].
\label{eq:Ibeta}
\end{multline}
Analytic continuation and substitution into Eq.~\eqref{eq:Kubo_sigmaV} then give
\begin{multline}
\sigma_V^{\mathrm{GUP}}(\mu,T,\beta)=I_\beta(\mu,T)
=\frac{1}{2\pi^2}\int_0^\infty dP\, P\, W_\beta(P)\times\\\left[n_F(P-\mu)+n_F(P+\mu)\right].
\label{eq:sigmaV_integral_GUP}
\end{multline}
{
Within the source convention adopted here, no additional mixed
\(A_i\)-\(h_{0j}\) contact vertex is included in the
\(J^iT^{0j}\) correlator. Possible source-dependent contributions
beyond the DoS sector are organized explicitly in
Sec.~\ref{sec:consistency_anomalies}.
}
Substituting \(\beta=0\) in Eq.~\eqref{eq:sigmaV_integral_GUP} gives
\begin{equation}
\sigma_V^{(0)}(\mu,T)
=
\frac{1}{2\pi^2}
\int_0^\infty dP\,P\,
\Big[n_F(P-\mu)+n_F(P+\mu)\Big].
\label{eq:sigmaV0_integral}
\end{equation}
Using
\begin{equation}
\int_0^\infty dP\,P\,n_F(P-\mu)
=
-T^2\operatorname{Li}_2(-e^{\mu/T}),
\label{eq:int_P_Li2}
\end{equation}
together with
\begin{equation}
\operatorname{Li}_2\!\bigl(-e^{a}\bigr)+\operatorname{Li}_2\!\bigl(-e^{-a}\bigr)
=
-\frac{a^2}{2}-\frac{\pi^2}{6},
\qquad
a\equiv \frac{\mu}{T},
\label{eq:Li2_identity}
\end{equation}
one obtains
\begin{equation}
\sigma_V^{(0)}(\mu,T)
=
\frac{\mu^2}{4\pi^2}
+
\frac{T^2}{12},
\label{eq:sigmaV0_result}
\end{equation}
which is the standard result for a single right-handed Weyl fermion of unit charge. For rigid rotation,
\begin{equation}
J^z_{(0)}
=
\sigma_V^{(0)}\,\omega^z
=
\Omega\left(
\frac{\mu^2}{4\pi^2}
+
\frac{T^2}{12}
\right).
\label{eq:Jz0}
\end{equation}
Using the weak-deformation expansion in Eq.~\eqref{eq:MeasureExpansion},
Eq.~\eqref{eq:sigmaV_integral_GUP} becomes
\begin{multline}
\sigma_V^{\rm GUP}(\mu,T,\beta)
=
\frac{1}{2\pi^2}
\int_0^\infty dP\,P\,
\Big[n_F(P-\mu)+n_F(P+\mu)\Big]
-\\\frac{3\beta}{2\pi^2}
\int_0^\infty dP\,P^3\,
\Big[n_F(P-\mu)+n_F(P+\mu)\Big]
+\mathcal O(\beta^2)
\\\equiv
\sigma_V^{(0)}(\mu,T)
+
\beta\,\sigma_V^{(1)}(\mu,T)
+
\mathcal O(\beta^2).
\label{eq:sigma_expand}
\end{multline}
Equivalently, in the moment notation of Sec.~\ref{sec2},
\begin{multline}
\sigma_V^{\rm GUP}(\mu,T,\beta)
=
I_1(\mu,T,\beta)+\bar I_1(\mu,T,\beta)
=\\
\bigl[I_1(\mu,T,0)+\bar I_1(\mu,T,0)\bigr]
-\\
3\beta\bigl[I_3(\mu,T,0)+\bar I_3(\mu,T,0)\bigr]
+
\mathcal O(\beta^2),
\label{eq:sigma_moment_form}
\end{multline}
which makes explicit that the \(O(\beta)\) correction is governed by the cubic thermal moment. The needed cubic moment is
\begin{equation}
\int_0^\infty dP\,P^3\,n_F(P-\mu)
=
-6T^4\operatorname{Li}_4(-e^{\mu/T}),
\label{eq:int_P3_Li4}
\end{equation}
and the relevant polylogarithmic identity is
\begin{equation}
\operatorname{Li}_4\!\bigl(-e^{a}\bigr)+\operatorname{Li}_4\!\bigl(-e^{-a}\bigr)
=
-\frac{a^4}{24}
-\frac{\pi^2 a^2}{12}
-\frac{7\pi^4}{360},
\label{eq:Li4_identity}
\end{equation}
where $a\equiv \mu/T.$ Therefore,
\begin{multline}
\int_0^\infty dP\,P^3
\Big[n_F(P-\mu)+n_F(P+\mu)\Big]
=\\
\frac{\mu^4}{4}
+
\frac{\pi^2}{2}\mu^2T^2
+
\frac{7\pi^4}{60}T^4.
\label{eq:P3_moment_result}
\end{multline}
Substituting Eq.~\eqref{eq:P3_moment_result} into Eq.~\eqref{eq:sigma_expand}, one finds
\begin{equation}
\sigma_V^{(1)}(\mu,T)
=
-\frac{3}{2\pi^2}
\left(
\frac{\mu^4}{4}
+
\frac{\pi^2}{2}\mu^2T^2
+
\frac{7\pi^4}{60}T^4
\right),
\label{eq:sigma1_result}
\end{equation}
so that
\begin{multline}
\sigma_V^{\rm GUP}(\mu,T,\beta)
=
\frac{\mu^2}{4\pi^2}
+
\frac{T^2}{12}
-\\
\beta\left[
\frac{3\mu^4}{8\pi^2}
+
\frac{3\mu^2T^2}{4}
+
\frac{7\pi^2}{40}T^4
\right]
+
\mathcal O(\beta^2)
\label{eq:sigmaV_GUP_boxed}
\end{multline}
for the present leading-order semiclassical GUP formalism. 
Eq.~\eqref{eq:sigmaV_GUP_boxed} is the explicit
\(\mathcal O(\beta)\) result for a single right-handed, unit-charge
Weyl fermion. Its undeformed part reproduces the standard chiral
vortical coefficient
\cite{Landsteiner:2011cp,Loganayagam:2012pz,Stone:2018zel},
whereas the minimal-length contribution forms a homogeneous
degree-four polynomial in \(\mu\) and \(T\). Dimensional analysis fixes
this thermodynamic degree, but not the numerical coefficients, their
relative sign, or their dependence on the chiral charge content; these
features follow from the parity-odd Kubo correlator. The result therefore
provides the single-species starting point for the two analyses that
follow: the unexpanded KMM integral establishes the behavior at finite
positive \(\beta\), while Sec.~\ref{sec:consistency_anomalies} resolves
the degree-four terms for a general chiral multiplet in terms of
\(d_{ABC}\), \(b_A\), and the fifth-order chiral charge trace
\(\mathcal D^{(5)}_{ABCDE}\)
\cite{Neiman:2010zi,Loganayagam:2012pz}. 
{ 
There is also a stronger statement that does not rely on expanding in
$\beta$.  From the exact integral representation in Eq.~\eqref{eq:sigmaV_integral_GUP}, the KMM
DoS contribution for the single right-handed
unit-charge Weyl fermion obeys
\begin{multline}
\frac{\partial \sigma_V^{\rm KMM}}{\partial\beta}
=
-\frac{3}{2\pi^2}
\int_0^\infty dP\,
\frac{P^3}
     {(1+\beta P^2)^4}\times\\
\left[
n_F(P-\mu)+n_F(P+\mu)
\right]
<0,
\label{eq:exact_monotonicity}
\end{multline}
where $\beta>0.$ Therefore the suppression of the DoS contribution is
exact for positive $\beta$ within the present framework and is not
merely a consequence of truncating the weak-deformation series.
Eq.~\eqref{eq:sigmaV_GUP_boxed} gives its leading expansion when
\(\beta\,\max\{\mu^2,T^2\}\ll1.\)
Outside this domain the unexpanded integral in Eq.~\eqref{eq:sigmaV_integral_GUP} should be used.} 
The leading minimal-length contribution is therefore
\begin{equation}
\delta_\beta\sigma_V
=
-\beta\left[
\frac{3\mu^4}{8\pi^2}
+\frac{3\mu^2T^2}{4}
+\frac{7\pi^2T^4}{40}
\right]
+\mathcal O(\beta^2),
\end{equation}
with no independent \(\mathcal O(\beta\mu^2)\) or
\(\mathcal O(\beta T^2)\) term. Thus, within the DoS
sector considered here, the deformation alters the higher thermal
moments while the coefficients of the standard degree-two
\(\mu^2\) and \(T^2\) contributions retain their undeformed
right-handed Weyl values. Their relation to the anomaly traces for a
general chiral multiplet is examined in
Sec.~\ref{sec:consistency_anomalies}. 
For rigid rotation, \(\omega^z=\Omega\), and the corresponding chiral
current is
\begin{equation}
J^z
=
\sigma_V^{\rm GUP}\,\omega^z
=
\Omega\,\sigma_V^{\rm GUP}(\mu,T,\beta),
\label{eq:Jz_final}
\end{equation}
where \(\sigma_V^{\rm GUP}\) is given in
Eq.~\eqref{eq:sigmaV_GUP_boxed}. 
{The resulting dependence on \(\mu\), \(T\), and \(\beta\) provides the basis for the dimensionless suppression analyzed in Sec.~\ref{sec:consistency_anomalies}. Its beam-energy dependence is considered separately in Sec.~\ref{sec:phenomenology}.}

\section{Consistency and Anomalies}
\label{sec:consistency_anomalies}
\noindent
Having obtained the single-species Kubo result, we now examine its
anomaly content and the chiral charge structure of the
\(\beta\)-dependent terms. Since the chirality, \(U(1)\) charge
assignment, current operator, and Belinfante stress-tensor vertex
remain those of the undeformed Weyl theory, the gauge and mixed
gauge-gravitational anomaly coefficients are inherited from the
underlying chiral theory. In anomalous hydrodynamics and hydrostatic
formulations, these anomaly data organize the leading vortical
contributions through the corresponding chiral charge traces
\cite{Neiman:2010zi,Loganayagam:2012pz,
Jensen:2013kka,Stone:2018zel}. 
The higher thermal structures obtained in Sec.~\ref{sec3} should
therefore be classified by their chiral charge content rather than by
their powers of \(\mu\) and \(T\) alone. This distinction allows one
to determine whether a given term carries the same charge trace as an
anomalous contribution while having a coefficient that is not fixed
by the anomaly polynomial
\cite{Loganayagam:2012pz,Jensen:2013kka,Stone:2018zel}.
For a single right-handed Weyl fermion of unit \(U(1)\) charge, the
covariant current obeys the standard gauge and mixed
gauge-gravitational anomaly relation
\cite{Landsteiner:2011cp,Bilal:2008qx,Stone:2018zel}.
In the covariant-current normalization used below, one may write
\begin{equation}
\nabla_\mu J^\mu_{\rm cov}
=
\frac{1}{32\pi^2}\,\epsilon^{\mu\nu\rho\sigma}F_{\mu\nu}F_{\rho\sigma}
+
\frac{1}{768\pi^2}\,
\epsilon^{\mu\nu\rho\sigma}
R^{\alpha}{}_{\beta\mu\nu}
R^{\beta}{}_{\alpha\rho\sigma},
\label{eq:anomaly_eq_cov}
\end{equation}
while the corresponding six-form anomaly polynomial can be written abstractly as
\begin{equation}
\begin{aligned}
\mathcal P_6
=
&\frac{\kappa_A}{6}\,F\wedge F\wedge F
+
\frac{\kappa_g}{24}\,F\wedge p_1(R),
\\
p_1(R)&\equiv -\frac{1}{8\pi^2}\,\mathrm{Tr}(R\wedge R).
\label{eq:anomaly_poly}
\end{aligned}
\end{equation}
The coefficients \(\kappa_A\) and \(\kappa_g\) are fixed by the
chirality and \(U(1)\) charge of the fermion
\cite{Loganayagam:2012pz,Jensen:2013kka}.
Eq.~\eqref{eq:anomaly_eq_cov} is written in the
covariant-current normalization, and this convention should be
distinguished from that used in the thermal Kubo derivation of
Sec.~\ref{sec3}
\cite{Landsteiner:2011cp,Landsteiner:2012kd,Jensen:2013kka}.
There, the chemical potential is implemented through the shifted
fermionic Matsubara frequencies of the free Weyl theory, and no
separate background Chern-Simons contribution is added to the mixed
\(J^iT^{0j}\) correlator. 
This distinction does not affect the DoS contribution
isolated here. The factor \(W_\beta(P)\) is a rotational scalar that
changes only the radial momentum integral, while the current vertex,
Belinfante stress-tensor vertex, chirality, and parity-odd tensor
structure remain those of the undeformed Weyl theory.  
In the normalization adopted in Sec.~\ref{sec3}, these anomaly data
reproduce the undeformed single-Weyl vortical conductivity derived in
Eq.~\eqref{eq:sigmaV0_result} \cite{Vilenkin:1979ui,Son:2009tf,Landsteiner:2011cp,Landsteiner:2016led}. 
{With the microscopic content fixed as specified in
Sec.~\ref{sec2}, the DoS contribution changes the
radial thermal coefficients without enlarging the parity-odd tensor
basis, as demonstrated explicitly in App.~\ref{A.2}. We now classify
the \(\beta\)-dependent terms according to their chiral charge
structure.}
This applies to the DoS contribution isolated
here and is not a general nonrenormalization statement for a
microscopic GUP-deformed chiral field theory. In a source-covariant
formulation, variations with respect to background gauge and
gravitational fields may also contribute through propagators, source
vertices, local contact terms, and the associated Ward identities. The latter constrain the complete correlator rather than the DoS contribution separately \cite{Banerjee:2012iz,Jensen:2013kka}.
{ It is useful to state explicitly how this contribution would appear
inside a source-covariant microscopic completion.  Suppose such a
theory admitted an expansion
\begin{equation}
S_\beta[\psi,A,g]
=
S_0[\psi,A,g]
+
\beta S_1[\psi,A,g]
+
\mathcal O(\beta^2),
\end{equation}
with
\begin{equation}
J_A^\mu
=
\frac{1}{\sqrt{-g}}
\frac{\delta S_\beta}{\delta A_\mu^A},
\qquad
T^{\rho\sigma}
=
\frac{2}{\sqrt{-g}}
\frac{\delta S_\beta}{\delta g_{\rho\sigma}}.
\end{equation}
The complete first-order correction to the mixed response would then
have the schematic decomposition
\begin{equation}
\delta_\beta G_{J_AT}
=
\delta G_{\rm DOS}
+
\delta G_{\rm prop}
+
\delta G_{J{\rm -vert}}
+
\delta G_{T{\rm -vert}}
+
\delta G_{\rm ct},
\label{eq:microscopic_decomposition}
\end{equation}
where $\delta G_{\rm DOS}$ is the DoS contribution
evaluated in this work, while the remaining terms denote,
respectively, corrections to the propagator, current vertex,
stress-tensor vertex, and mixed contact contribution.  In particular,
a source-dependent microscopic action could contain
\begin{equation}
\delta G_{\rm ct}
\supset
\frac{\delta^2 S_1}
{\delta A_\mu^A\,\delta g_{\rho\sigma}}.
\end{equation}
The Ward identities of a source-covariant completion would constrain
the sum in Eq.~\eqref{eq:microscopic_decomposition}, not
$\delta G_{\rm DOS}$ separately.  The present calculation therefore
determines one well-defined contribution to that complete response
and does not assume that the other terms vanish in an arbitrary GUP
field theory.
}
{ 
Starting from the single-Weyl result derived in
Sec.~\ref{sec3}, we now extend the vortical conductivity to a
general massless chiral multiplet in order to resolve the charge
traces carried by the higher thermal terms.
Let $f$ label Weyl
species, let
\begin{equation}
\chi_f=
\begin{cases}
+1, & f\ {\rm right\mbox{-}handed},\\
-1, & f\ {\rm left\mbox{-}handed},
\end{cases}
\end{equation}
and let $q_A^f$ denote the charge of species $f$ under a mutually
commuting conserved current $J_A^\mu$.  For chemical potentials
$\mu_B$ associated with these charges,
\(\mu_f=q_B^f\mu_B .\)
Assuming the same isotropic minimal-length factor for all species,
the DoS contribution generalizing Eq.~\eqref{eq:sigmaV_integral_GUP} is
\begin{multline}
\sigma^{\rm DOS}_{V,A}
=
\frac{1}{2\pi^2}
\sum_f\chi_f q_A^f
\int_0^\infty dP\,P\,W_\beta(P)\times \\
\left[
n_F(P-\mu_f)+n_F(P+\mu_f)
\right].
\label{eq:general_chiral_kubo}
\end{multline}
{
To make the generalization stated above explicit, consider an analytic
rotationally invariant density-of-states deformation of the form
\begin{equation}
W_\beta(P)
=
w(\beta P^2)
=
1+c_1\beta P^2+\mathcal{O}(\beta^2),
\label{eq:general_isotropic_weight}
\end{equation}
where $w(0)=1, \,\, c_1=w'(0)$. At first order in $\beta$, and within the density-of-states sector
considered here, the dependence on the particular analytic isotropic
deformation is therefore carried by the coefficient $c_1$.
Rotational invariance ensures that the deformation changes the radial
thermal moments without enlarging the parity-odd tensor basis of the
mixed $J^iT^{0j}$ correlator, as established explicitly in
App.~\ref{App:a}. For the KMM density of states,
$w(z)=(1+z)^{-3}$, so that $c_1=-3$
\cite{Kempf:1994su,Chang:2001bm,Fityo:2008zz}. The first-order
charge-trace decomposition derived below consequently applies to this
general analytic isotropic DoS class, whereas the exact
finite-$\beta$ monotonicity result remains specific to the KMM form.} 
The anomaly traces in the
conventions of Refs.~\cite{Landsteiner:2011cp,Amado:2011zx,Landsteiner:2012kd} are
\begin{equation}
d_{ABC}
=
\frac{1}{2}
\left[
{\rm tr}_{R}\!\left(T_A\{T_B,T_C\}\right)
-
{\rm tr}_{L}\!\left(T_A\{T_B,T_C\}\right)
\right],
\label{eq:anomaly_traces}
\end{equation}
where $b_A ={\rm tr}_{R}(T_A)-{\rm tr}_{L}(T_A)$. For mutually commuting charges these reduce to
\begin{equation}
d_{ABC}
=
\sum_f
\chi_f q_A^f q_B^f q_C^f,
\,\,
b_A
=
\sum_f
\chi_f q_A^f .
\end{equation}
We also define the fifth-order chiral charge trace
\begin{equation}
{\cal D}^{(5)}_{ABCDE}
\equiv
\sum_f
\chi_f
q_A^f q_B^f q_C^f q_D^f q_E^f .
\label{eq:fifth_trace}
\end{equation}
Using
\begin{equation}
\int_0^\infty dP\,P
\left[n_F(P-\mu)+n_F(P+\mu)\right]
=
\frac{\mu^2}{2}
+
\frac{\pi^2T^2}{6},
\end{equation}
and
\begin{multline}
\int_0^\infty dP\,P^3
\left[n_F(P-\mu)+n_F(P+\mu)\right]
=\\
\frac{\mu^4}{4}
+
\frac{\pi^2}{2}\mu^2T^2
+
\frac{7\pi^4}{60}T^4,
\end{multline}
Eq.~\eqref{eq:general_chiral_kubo} becomes
\begin{widetext}
\begin{equation}
\sigma^{\rm DOS}_{V,A}
={}
\frac{1}{4\pi^2}
d_{ABC}\mu_B\mu_C
+
\frac{T^2}{12}b_A
+
c_1\beta
\left[
\frac{1}{8\pi^2}
{\cal D}^{(5)}_{ABCDE}
\mu_B\mu_C\mu_D\mu_E
+
\frac{1}{4}
d_{ABC}\mu_B\mu_C T^2
+
\frac{7\pi^2}{120}b_A T^4
\right]
+
\mathcal O(\beta^2).
\label{eq:general_trace_result}
\end{equation}
\end{widetext}
For the KMM DoS, $c_1=-3$, and therefore
\begin{multline}
\delta_\beta\sigma^{\rm DOS}_{V,A}
={}
-\beta
\Bigg[
\frac{3}{8\pi^2}
{\cal D}^{(5)}_{ABCDE}
\mu_B\mu_C\mu_D\mu_E
+\\
\frac{3}{4}
d_{ABC}\mu_B\mu_C T^2
+
\frac{7\pi^2}{40}b_A T^4
\Bigg]
+
\mathcal O(\beta^2).
\label{eq:KMM_trace_result}
\end{multline}
Eqs.~\eqref{eq:general_trace_result} and
\eqref{eq:KMM_trace_result} display explicitly the chiral charge-trace
structure of the deformation-dependent terms. 
The deformation-dependent $\mu_B\mu_C T^2$ term carries the same
$d_{ABC}$ trace that occurs in the gauge anomaly, while the $T^4$
term carries the same $b_A$ trace associated with the mixed
gauge-gravitational anomaly.  The purely quartic
chemical-potential contribution instead contains the independent
fifth-order trace ${\cal D}^{(5)}_{ABCDE}$.  Consequently, the
higher-order terms should not be described as entirely disconnected
from anomalous structures.  The sharper statement is that they are
not anomaly-fixed: their coefficients depend explicitly on the
DoS function through $c_1$, and the fifth-order trace is
not one of the coefficients appearing in the four-dimensional anomaly
polynomial. 
For a single right-handed unit-charge Weyl fermion,
\(d=b=\mathcal D^{(5)}=1\), and
Eq.~\eqref{eq:KMM_trace_result} reproduces the single-species
first-order contribution derived in Sec.~\ref{sec3}.
The absence of independent degree-two
\(\mathcal O(\beta)\) terms follows from the same result.} 
{The exact monotonicity result established in
Sec.~\ref{sec3} implies a positive suppression for \(\beta>0\).
We therefore define}
\begin{equation}
\Delta_\beta(x)
\equiv
-
\frac{\delta\sigma_V}{\sigma_V^{(0)}}
=
-
\frac{\sigma_V^{\rm GUP}-\sigma_V^{(0)}}{\sigma_V^{(0)}},
\qquad
x\equiv \frac{\mu}{T},
\label{eq:Delta_beta_def}
\end{equation}
which to leading order becomes
\begin{equation}
\Delta_\beta(x)
=
(\beta T^2)\,
\frac{
\dfrac{3x^4}{8\pi^2}
+
\dfrac{3x^2}{4}
+
\dfrac{7\pi^2}{40}
}{
\dfrac{x^2}{4\pi^2}
+
\dfrac{1}{12}
}
+
\mathcal O\!\big((\beta T^2)^2\big).
\label{eq:Delta_beta_x}
\end{equation}
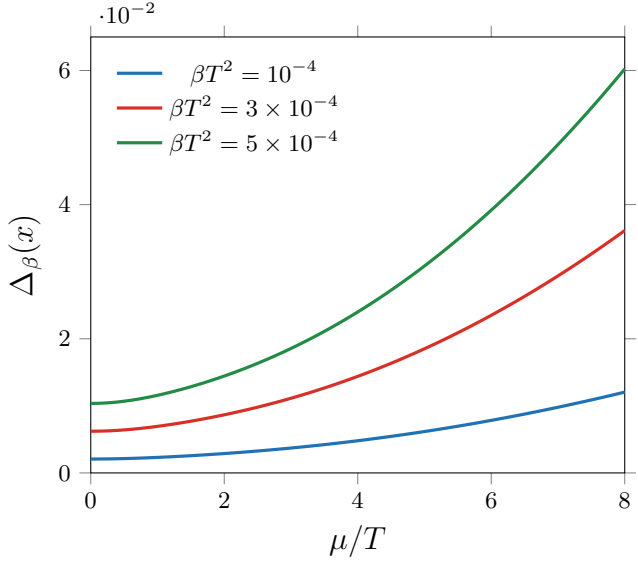
\begin{figure}[htb]
\centering
\begin{tikzpicture}
\begin{axis}[
  width=\columnwidth,
  height=0.85\columnwidth,
  xlabel={$\mu/T$},
  ylabel={$\Delta_\beta(x)$},
  xmin=0, xmax=8,
  ymin=0, ymax=0.065,
  domain=0:8,
  samples=240,
  tick align=outside,
  grid=none,
  major grid style={draw=gray!20},
  legend style={at={(0.03,0.97)},anchor=north west,draw=none,fill=none,font=\small},
  label style={font=\large},
  declare function={
    dfrac(\x,\b)=\b*((3*\x^4)/(8*pi^2) + (3*\x^2)/4 + (7*pi^2)/40)/((\x^2)/(4*pi^2) + 1/12);
  },
]
\addplot[very thick,myblue] {dfrac(x,1e-4)};
\addlegendentry{$\beta T^2=10^{-4}$}

\addplot[very thick,myred] {dfrac(x,3e-4)};
\addlegendentry{$\beta T^2=3\times10^{-4}$}

\addplot[very thick,mygreen] {dfrac(x,5e-4)};
\addlegendentry{$\beta T^2=5\times10^{-4}$}
\end{axis}
\end{tikzpicture}
\caption{\justifying Fractional suppression \(\Delta_\beta\) of the chiral vortical
conductivity as a function of \(\mu/T\) for three representative
values of \(\beta T^2\). For \(\beta>0\), the suppression increases
with \(\mu/T\).}
\label{fig:delta_mu_over_T}
\end{figure}
Fig.~\ref{fig:delta_mu_over_T} displays this quantity as a function of \(\mu/T\) at fixed
\(\beta T^2\). The beam-energy dependence of \(\Delta_\beta\) and its relation to
chemical freeze-out conditions are considered in
Sec.~\ref{sec:phenomenology}.
\section{Phenomenological Implications for Heavy-Ion Collisions}
\label{sec:phenomenology}

\noindent
{ We now turn to the heavy-ion interpretation of the equilibrium vortical conductivity derived above. Global hyperon-polarization measurements establish the presence of a vortical medium but do not provide a direct determination of \(\sigma_V\) 
\cite{STAR:2017ckg,STAR:2018gyt,STAR:2023nvo,STAR:2021beb,HADES:2022enx,ALICE:2019onw}. 
The vector and axial vortical responses must be distinguished when the medium possesses independent vector and axial chemical potentials \cite{Vilenkin:1979ui,Son:2009tf}.  For one massless Dirac fermion define
\begin{equation}
\mu_R=\mu_V+\mu_A,
\qquad
\mu_L=\mu_V-\mu_A,
\end{equation}
and introduce the single-right-handed response
\begin{equation}
F_\beta(\mu,T)
\equiv
\frac{1}{2\pi^2}
\int_0^\infty dP\,P\,W_\beta(P)
\left[
n_F(P-\mu)+n_F(P+\mu)
\right].
\end{equation}
The vector and axial vortical conductivities are then
\begin{equation}
\begin{aligned}
\sigma_{V,\mathrm{vec}}
=
F_\beta(\mu_R,T)-F_\beta(\mu_L,T),
\\
\sigma_{V,\mathrm{ax}}
=
F_\beta(\mu_R,T)+F_\beta(\mu_L,T).
\label{eq:vector_axial_exact}
\end{aligned}
\end{equation}
Because $F_\beta(\mu,T)$ is even in $\mu$, Eq.~\eqref{eq:vector_axial_exact}
implies exactly
\begin{equation}
\mu_A=0
\quad\Longrightarrow\quad
\sigma_{V,\mathrm{vec}}=0,
\quad
\sigma_{V,\mathrm{ax}}=2F_\beta(\mu_V,T)\neq0
\end{equation}
for a nontrivial thermal or finite-density state.  In the undeformed
limit this reduces, with the vorticity convention used in this work,
to
\begin{equation}
\sigma^{(0)}_{V,\mathrm{vec}}
=
\frac{\mu_V\mu_A}{\pi^2},
\,\,
\sigma^{(0)}_{V,\mathrm{ax}}
=
\frac{\mu_V^2+\mu_A^2}{2\pi^2}
+
\frac{T^2}{6}.
\label{eq:vector_axial_standard}
\end{equation}
Thus a vortical vector current requires a nonzero axial imbalance,
whereas the axial vortical current remains finite at $\mu_A=0$ through
the $\mu_V^2$ and $T^2$ contributions
\cite{Landsteiner:2011cp,Landsteiner:2016led}. 
A second distinction concerns the observable itself.  The axial
current is the local coordinate-space expectation value
\begin{equation}
J_A^\mu(x)
=
\left\langle
\bar{\psi}(x)\gamma^\mu\gamma^5\psi(x)
\right\rangle ,
\end{equation}
whereas hyperon polarization is a momentum-dependent spin observable,
which may be represented by a spin four-vector
$S_\Lambda^\mu(p)$ obtained from the spin density matrix or Wigner
function and subsequently integrated over the freeze-out hypersurface
\cite{Becattini:2013fla,Palermo:2021hlf}.  These quantities respond to the
same rotating medium but are not interchangeable:
\begin{equation}
S_\Lambda^\mu(p)
\not\equiv
C\,J_A^\mu(x),
\qquad
P_\Lambda
\not\equiv
C'\,\sigma_V
\end{equation}
for universal constants $C$ or $C'$.  
A quantitative relation between these quantities requires the
spacetime evolution, hadronization, feed-down, and freeze-out
information absent from the equilibrium Kubo coefficient.
Accordingly, the analysis below uses chemical freeze-out relations
only to evaluate the equilibrium conductivity ratio and does not
identify it with a measured hadronic current.}

For phenomenological orientation, we relate the beam energy
\(\sqrt{s_{NN}}\) to the chemical freeze-out parameters using the
standard thermal-fit parametrization \cite{Cleymans:2005xv,Mir:2023wkm}
\begin{equation}
\mu_B(\sqrt{s_{NN}})
=
\frac{d}{1+e\,\sqrt{s_{NN}}},
\label{eq:muB_of_s}
\end{equation}
together with
\begin{equation}
T(\mu_B)
=
a-b\,\mu_B^2-c\,\mu_B^4,
\label{eq:T_of_muB}
\end{equation}
where the parameters are as follows: $d = 1.308~{\rm GeV},\,e = 0.273~{\rm GeV}^{-1},\,a=0.166~{\rm GeV},\,b=0.139~{\rm GeV}^{-1},\,c=0.053~{\rm GeV}^{-3}.$
Fig.~\ref{fig:freezeout_phaseplane} shows the corresponding freeze-out trajectory in the \((\mu_B,T)\) plane with representative beam energies marked explicitly. The main trend is the familiar one: toward lower \(\sqrt{s_{NN}}\), the baryon chemical potential
increases substantially, whereas the freeze-out temperature varies
more moderately. This is the region in which vortical phenomena are
expected to be most relevant phenomenologically.
\begin{widetext}
\begin{figure*}[htb]
\centering
\begin{tikzpicture}
\begin{axis}[
  width=0.70\textwidth,
  height=0.65\textwidth,
  xlabel={$\mu_B$ (GeV)},
  ylabel={$T$ (GeV)},
  xmin=0, xmax=0.78,
  ymin=0.075, ymax=0.170,
  tick align=outside,
  grid=none,
  major grid style={draw=gray!20},
  minor grid style={draw=gray!10},
  tick label style={font=\small},
  label style={font=\large},
  every axis plot/.append style={line width=1.0pt},
]
\addplot[
  very thick,
  blue!70!black,
  domain=0:0.75,
  samples=240
]
{0.166 - 0.139*x^2 - 0.053*x^4};
\addplot+[
  only marks,
  mark=*,
  mark size=2.6pt,
  red!75!black
]
coordinates {
  (0.7191,0.0800)   
  (0.4216,0.1396)   
  (0.2060,0.1600)   
  (0.1563,0.1626)   
  (0.0235,0.1659)   
  (0.0017,0.1660)   
};
\node[font=\scriptsize,anchor=north east] at (axis cs:0.7191,0.0800) {$3$ GeV};
\node[font=\scriptsize,anchor=south west] at (axis cs:0.4216,0.1396) {$7.7$ GeV};
\node[font=\scriptsize,anchor=south east] at (axis cs:0.3060,0.1600) {$19.6$ GeV};
\node[font=\scriptsize,anchor=south east] at (axis cs:0.2063,0.1626) {$27$ GeV};
\node[font=\scriptsize,anchor=south west] at (axis cs:0.0235,0.1659) {$200$ GeV};
\node[font=\scriptsize,anchor=south west] at (axis cs:0.0017,0.1600) {LHC};
\node[
  font=\scriptsize,
  fill=none,
  inner sep=2pt,
  anchor=south west
] at (axis cs:0.40,0.090)
{chemical freeze-out trajectory};
\end{axis}
\end{tikzpicture}
\caption{\justifying chemical freeze-out trajectory in the \((\mu_B,T)\) plane obtained from the standard parametrization of Eq.~\eqref{eq:T_of_muB}. The marked points indicate representative beam energies relevant for fixed-target, RHIC, and the low-\(\mu_B\) LHC end of the trajectory. The figure makes explicit that lower \(\sqrt{s_{\mathrm{NN}}}\) corresponds to substantially larger baryon chemical potential, while the freeze-out temperature changes more moderately.}
\label{fig:freezeout_phaseplane}
\end{figure*}
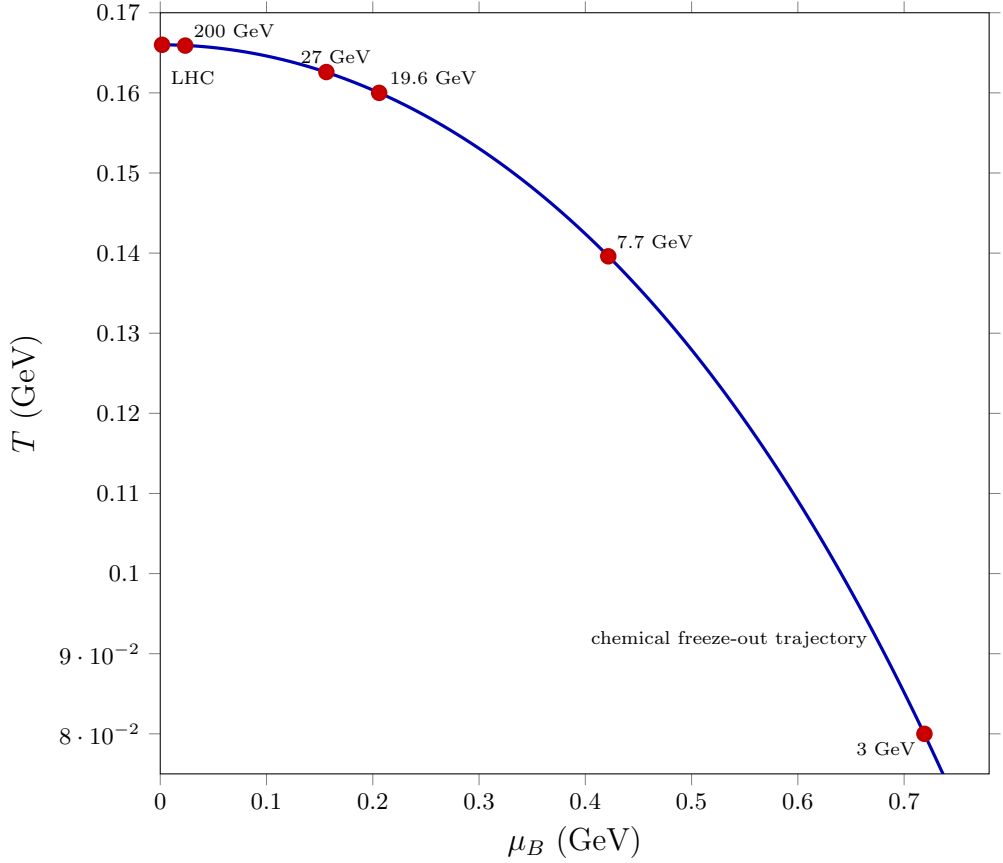
\end{widetext}
{ To evaluate the single-species result along the freeze-out trajectory,
we use the illustrative quark-level identification \cite{Mir:2023wkm,Mir:2024wlo,Mir:2025qqv}
\begin{equation}
\mu_q(\sqrt{s_{NN}})
=
\frac{\mu_B(\sqrt{s_{NN}})}{3},
\,\,
x_q(\sqrt{s_{NN}})
\equiv
\frac{\mu_q(\sqrt{s_{NN}})}
     {T(\sqrt{s_{NN}})} .
\label{eq:muq_freezeout}
\end{equation}
This identification serves only to evaluate the equilibrium
single-species result of Sec.~\ref{sec3} along the freeze-out
trajectory. A baryon, electric, or axial current in QCD requires the
corresponding flavor, charge, and chirality sums. 
For the beam-energy analysis, we take
\(T_{\rm ref}=0.16~{\rm GeV}\) and express the deformation through
the dimensionless combination \(\beta T_{\rm ref}^{2}\).
Substituting \(T(\sqrt{s_{NN}})\) and
\(x_q(\sqrt{s_{NN}})\) into
Eq.~\eqref{eq:Delta_beta_x}, the fractional suppression becomes
\begin{widetext}
\begin{equation}
\Delta_\beta(\sqrt{s_{NN}})
=
(\beta T_{\rm ref}^{2})
\left[
\frac{T(\sqrt{s_{NN}})}{T_{\rm ref}}
\right]^2\times
\frac{
\displaystyle
\frac{3x_q^4(\sqrt{s_{NN}})}{8\pi^2}
+
\frac{3x_q^2(\sqrt{s_{NN}})}{4}
+
\frac{7\pi^2}{40}
}{
\displaystyle
\frac{x_q^2(\sqrt{s_{NN}})}{4\pi^2}
+
\frac{1}{12}
}
+
\mathcal O\!\left[
(\beta T_{\rm ref}^{2})^2
\right].
\label{eq:Delta_beta_beam}
\end{equation}
\end{widetext}
}
{
The beam-energy dependence in Eq.~\eqref{eq:Delta_beta_beam} is shown in Fig.~\ref{fig:beam_energy_suppression}. Its variation reflects the combined dependence of \(T(\sqrt{s_{NN}})\) and \(\mu_q(\sqrt{s_{NN}})\) along the chemical freeze-out trajectory. No identification of \(\Delta_\beta\) with a measured hyperon-polarization observable is made here.
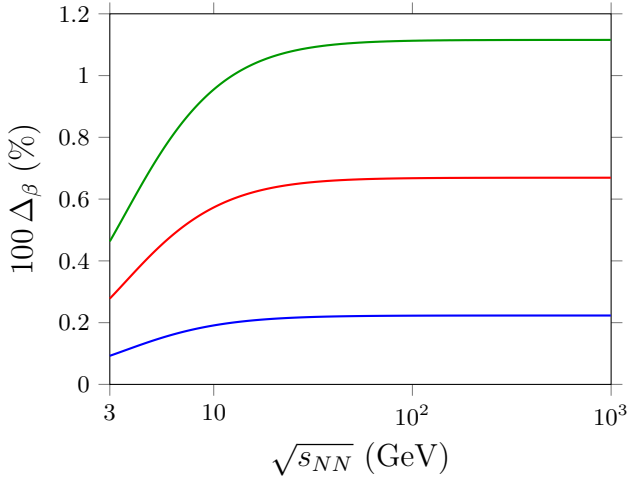
\begin{figure}[htb]
\centering
\begin{tikzpicture}
\begin{axis}[
    width=0.95\columnwidth,
    height=0.75\columnwidth,
    xmode=log,
    xmin=3,
    xmax=1000,
    ymin=0,
    ymax=1.2,
    xlabel={$\sqrt{s_{NN}}~(\mathrm{GeV})$},
    ylabel={$100\,\Delta_{\beta}\;(\%)$},
    xtick={3,10,100,1000},
    xticklabels={$3$,$10$,$10^{2}$,$10^{3}$},
    ytick={0,0.2,0.4,0.6,0.8,1.0,1.2},
    label style={font=\large},
    legend style={
        draw=none,
        fill=none,
        at={(0.96,0.93)},
        anchor=north east,
        font=\small
    },
    tick align=outside,
    axis line style={thin},
    tick style={thin},
    samples=250,
    domain=3:1000,
    declare function={
        muB(\s)=1.308/(1+0.273*\s);
        Tfo(\s)=0.166
                -0.139*muB(\s)^2
                -0.053*muB(\s)^4;
        xq(\s)=(muB(\s)/3)/Tfo(\s);
        Rfun(\s)=(
            3*xq(\s)^4/(8*pi^2)
            +3*xq(\s)^2/4
            +7*pi^2/40
        )/(
            xq(\s)^2/(4*pi^2)
            +1/12
        );
    }
]

\addplot[
    blue,
    thick
]
{
    100*(1e-4)
    *(Tfo(x)/0.16)^2
    *Rfun(x)
};

\addplot[
    red,
    thick
]
{
    100*(3e-4)
    *(Tfo(x)/0.16)^2
    *Rfun(x)
};

\addplot[
    green!60!black,
    thick
]
{
    100*(5e-4)
    *(Tfo(x)/0.16)^2
    *Rfun(x)
};

\end{axis}
\end{tikzpicture}

\caption{\justifying
Beam-energy dependence of the fractional DoS
suppression $\Delta_\beta$ of the equilibrium chiral vortical
conductivity along the chemical freeze-out trajectory, using the
illustrative relation $\mu_q=\mu_B/3$.
The curves correspond to
$\beta T_{\rm ref}^{2}=10^{-4}$ (green colored curve),
$3\times10^{-4}$ (red colored curve), and $5\times10^{-4}$ (blue colored curve), with
$T_{\rm ref}=0.16~\mathrm{GeV}$.
No hyperon-polarization data or other hadronic spin observable are
included.
}
\label{fig:beam_energy_suppression}
\end{figure}
}
{ In the freeze-out region shown in
Fig.~\ref{fig:freezeout_phaseplane}, both \(T\) and
\(|\mu_q|\) remain well below \(1~{\rm GeV}\).
To characterize the hierarchy relevant to the minimal-length
contribution, we define
\begin{equation}
\epsilon_{\rm ML}
\equiv
\beta\Lambda^2,
\qquad
\Lambda
\equiv
\max\{T,|\mu_q|\}.
\label{eq:epsilon_ML}
\end{equation}
Here
\(
\beta
=
\frac{\beta_0}{M_\star^2},
\)
therefore one obtains
\(
\epsilon_{\rm ML}
=
\beta_0
\left(
\frac{\Lambda}{M_\star}
\right)^2 .
\)
The relative DoS contribution to the vortical
conductivity is therefore parametrically of order
\(\epsilon_{\rm ML}\), up to numerical coefficients of order unity.
For a Planck-motivated deformation with
\(M_\star\sim M_{\rm Pl}\) and \(\beta_0=\mathcal O(1)\),
this contribution is negligible under heavy-ion freeze-out
conditions. 
For \(T,|\mu_q|<1~{\rm GeV}\) and
\(M_\star\sim10^{19}~{\rm GeV}\), this ratio is of order
\(10^{-38}\beta_0\) or smaller. The heavy-ion comparison therefore
demonstrates the enormous hierarchy between hadronic thermodynamic
energies and \(M_\star\).  It is not used to infer
hyperon polarization, to convert $\sigma_V$ into a measured spin
observable, or to place an experimental bound on $\beta_0$.}


\noindent
\section{Conclusion}
\label{sec6}
{We have derived the DoS contribution to chiral vortical
transport for massless chiral fermions and established its extension
from a single right-handed Weyl species to general chiral multiplets
with mutually commuting charges. The parity-odd Kubo analysis shows
that an isotropic DoS deformation modifies the radial
thermal moments without enlarging the vortical tensor structure. For
the analytic class
$W_\beta(P)=w(\beta P^2)$, the conventional degree-two
$\mu^2$ and $T^2$ contributions retain their undeformed coefficients
through first order in $\beta$ within this sector, whereas the leading
minimal-length correction is of degree four in the chemical potentials
and temperature. Its mixed chemical-potential thermal and purely
thermal terms carry the chiral traces associated with the gauge and
mixed gauge-gravitational anomaly structures, while the purely
quartic chemical-potential contribution contains an independent
fifth-order trace. The coefficients of these higher-order terms depend
on the DoS deformation and are therefore not fixed by
the four-dimensional anomaly polynomial.
For the KMM form, the exact integral representation further proves
that the DoS contribution to the vortical conductivity
decreases monotonically for positive $\beta$, so the perturbative
suppression is not an artifact of the first-order expansion. The
heavy-ion comparison shows that a Planck-suppressed deformation is
negligible along standard chemical freeze-out trajectories. The result
therefore provides a general characterization of isotropic
minimal-length phase-space corrections to equilibrium vortical
transport while keeping separate the additional microscopic
source-dependent contributions that may arise in a complete
GUP-deformed chiral field theory.}
\vspace{-0.2cm}
\section*{Acknowledgments}
\vspace{-0.2cm}
The author SAM is thankful to Mir Faizal of Durham University for fruitful discussions and understanding of many aspects of the Quantum Gravity. The author Euaggelos E. Zotos is thankful to Princess Nourah bint Abdulrahman University, Riyadh, Saudi Arabia for supporting the research via Project number (PNURSP2026R924).
\vspace{-0.4cm}
\section*{Data Availability Statement}
\vspace{-0.3cm}
The data supporting the findings of this study are included in the article and are referenced appropriately throughout the text. Additional results from this study will be made available upon request to the corresponding author.

\bibliographystyle{apsrev4-1}
\bibliography{refs_rf_gup-2}


\appendix

\onecolumngrid


\section{Explicit evaluation of the mixed $JT^{0j}$ Kubo correlator}
\label{App:a}
This appendix provides the detailed one-loop derivation of the mixed
\(J^iT^{0j}\) correlator used in Sec.~\ref{sec3}. The fermionic
sector and DoS factor are those defined in
Sec.~\ref{sec2}. App.~\ref{A.1} decomposes the Belinfante
momentum-density vertex, App.~\ref{A.2} isolates the parity-odd
rotational structure, and App.~\ref{A.3} performs the Matsubara sums,
vacuum subtraction, analytic continuation, and static
long-wavelength limit.

\subsection{Decomposition of the mixed correlator}
\label{A.1}
We next decompose the mixed Euclidean current-momentum density
correlator according to the two Dirac structures carried by the
Belinfante momentum-density vertex before taking the static
parity-odd limit. The chiral projector fixes the current insertion,
whereas \(W_\beta(P)\) appears as a rotational scalar multiplying the
spatial momentum integral. This separation allows the Dirac trace,
angular reduction, and Matsubara sum to be carried out independently
of the radial DoS factor \cite{Amado:2011zx,Landsteiner:2011cp,Landsteiner:2012kd}. We begin from the Euclidean two-point function
{\begin{equation}
G_E^{ij}(K)
=
\int_{0}^{1/T} d\tau
\int d^3x\,
e^{-i\omega_m\tau+i\mathbf{k}\cdot\mathbf{x}}
\left\langle
T_\tau J^i(\tau,\mathbf{x})T^{0j}(0,\mathbf{0})
\right\rangle,
\qquad
K=(i\omega_m,\mathbf{k}).
\end{equation}
Here $\omega_m=2\pi mT$ is a bosonic Matsubara frequency. The chemical potential is incorporated through the shifted fermionic frequency $i\widetilde{\omega}_n=i(2n+1)\pi T+\mu$. Accordingly, the internal fermionic four-momentum is written as $Q=(i\widetilde{\omega}_n,\mathbf{P})$, while $Q+K=(i\widetilde{\omega}_n+i\omega_m,\mathbf{P}+\mathbf{k})$.}
After Wick contraction, translational invariance, and insertion of the GUP weight into the spatial loop measure, the one-loop Euclidean correlator takes the form
\begin{equation}
\mathcal{G}_{E}^{\,ij}(K)
=
T\sum_{n\in\mathbb{Z}}
\int_{\mathbf{P}}
W_{\beta}(P)\,
\mathrm{tr}\!\left[
\gamma^{i}P_{R}\,S(Q)\,V^{0j}(Q,Q+K)\,S(Q+K)
\right],
\quad
\int_{\mathbf{P}}\equiv \int \frac{d^{3}P}{(2\pi)^{3}},
\end{equation}
where \(W_\beta(P)\) is defined in
Eq.~\eqref{eq:LiouvilleMeasure}, while the propagator and vertices
retain their undeformed free-theory forms. For the massless fermion one may write the propagator in the helicity decomposition
\begin{equation}
S(Q)
=
\frac{1}{2}\sum_{s=\pm}
\Delta_{s}(Q)\,\gamma_{\mu}\,\hat{P}_{s}^{\mu},
\quad
\Delta_{s}(Q)=\frac{1}{i\tilde{\omega}_{n}-sE_{P}},
\quad
E_{P}=P,
\quad
\hat{P}_{s}^{\mu}=(1,s\hat{\mathbf{P}}),
\end{equation}
and similarly
\begin{multline}
S(Q+K)
=
\frac{1}{2}\sum_{t=\pm}
\Delta_{t}(Q+K)\,\gamma_{\nu}\,\hat{R}_{t}^{\nu},
\quad
\Delta_{t}(Q+K)=\frac{1}{i\tilde{\omega}_{n}+i\omega_{m}-tE_{R}},
\\
E_{R}=|\mathbf{P}+\mathbf{k}|,
\qquad
\hat{R}_{t}^{\mu}=(1,t\widehat{\mathbf{P}+\mathbf{k}}).
\end{multline}
The momentum-density insertion is represented by the Belinfante vertex
\begin{equation}
V^{0j}(Q,Q+K)
=
\frac{1}{4}
\Bigl[
\gamma^{0}(2P+k)^{j}
+
\gamma^{j}\bigl(2i\tilde{\omega}_{n}+i\omega_{m}\bigr)
\Bigr]P_{R}.
\end{equation}
At this stage the correlator is exact within the one-loop
DoS formulation. To extract the vortical conductivity we now choose the external spatial momentum along the \(z\)-axis,
\begin{equation}
\mathbf{k}=k\,\hat{\mathbf{z}},
\qquad
K=(i\omega_{m},0,0,k),
\end{equation}
and define the parity-odd projected scalar correlator
\begin{equation}
\mathcal{G}_{V,E}(K)\equiv \frac{1}{2}\,\epsilon_{ij3}\,\mathcal{G}_{E}^{\,ij}(K).
\end{equation}
With this choice the projector immediately removes the explicit \(k^{j}\) term in the stress-tensor vertex, because
\begin{equation}
\epsilon_{ij3}\,k^{j}
=
k\,\epsilon_{ij3}\,\delta^{j3}
=
0.
\end{equation}
Furthermore, the term proportional to \(i\omega_{m}\gamma^{j}\) does not contribute to the static Kubo limit, since the vortical conductivity is extracted at \(\omega_{m}\to 0\) before or together with the long-wavelength expansion. Therefore only the reduced vertex
\begin{equation}
V_{\mathrm{red}}^{\,0j}(Q)
=
\frac{1}{2}\Bigl(\gamma^{0}P^{j}+\gamma^{j}i\tilde{\omega}_{n}\Bigr)P_{R}
\end{equation}
survives in the parity-odd projected channel relevant for \(\sigma_{V}\). Substituting this reduced vertex into the projected correlator gives
\begin{equation}
\mathcal{G}_{V,E}(K)
=
\frac{T}{4}
\sum_{n\in\mathbb{Z}}
\int_{\mathbf{P}}
W_{\beta}(P)\,
\epsilon_{ij3}\,
\mathrm{tr}\!\left[
\gamma^{i}P_{R}\,S(Q)\,
\Bigl(\gamma^{0}P^{j}+\gamma^{j}i\tilde{\omega}_{n}\Bigr)P_{R}\,
S(Q+K)
\right].
\end{equation}
This is the point at which the decomposition becomes both natural and unique:
\begin{equation}
\mathcal{G}_{V,E}(K)
=
\mathcal{G}_{V,E}^{(0j)}(K)
+
\mathcal{G}_{V,E}^{(j0)}(K),
\end{equation}
with
\begin{equation}
\mathcal{G}_{V,E}^{(0j)}(K)
=
\frac{T}{4}
\sum_{n\in\mathbb{Z}}
\int_{\mathbf{P}}
W_{\beta}(P)\,
\epsilon_{ij3}\,P^{j}\,
\mathrm{tr}\!\left[
\gamma^{i}P_{R}\,S(Q)\,\gamma^{0}P_{R}\,S(Q+K)
\right],
\end{equation}
and
\begin{equation}
\mathcal{G}_{V,E}^{(j0)}(K)
=
\frac{T}{4}
\sum_{n\in\mathbb{Z}}
\int_{\mathbf{P}}
W_{\beta}(P)\,
\epsilon_{ij3}\,i\tilde{\omega}_{n}\,
\mathrm{tr}\!\left[
\gamma^{i}P_{R}\,S(Q)\,\gamma^{j}P_{R}\,S(Q+K)
\right].
\end{equation}
Equivalently, after inserting the helicity decomposition of both propagators, one obtains the completely explicit forms
\begin{equation}
\mathcal{G}_{V,E}^{(0j)}(K)
=
\frac{T}{16}
\sum_{n\in\mathbb{Z}}
\int_{\mathbf{P}}
W_{\beta}(P)\,
\epsilon_{ij3}\,P^{j}
\sum_{s,t=\pm}
\Delta_{s}(Q)\Delta_{t}(Q+K)\,
\mathrm{tr}\!\left[
\gamma_{\mu}\gamma^{i}\gamma_{\nu}\gamma^{0}P_{R}
\right]
\hat{P}_{s}^{\mu}\hat{R}_{t}^{\nu},
\end{equation}
\begin{equation}
\mathcal{G}_{V,E}^{(j0)}(K)
=
\frac{T}{16}
\sum_{n\in\mathbb{Z}}
\int_{\mathbf{P}}
W_{\beta}(P)\,
\epsilon_{ij3}\,i\tilde{\omega}_{n}
\sum_{s,t=\pm}
\Delta_{s}(Q)\Delta_{t}(Q+K)\,
\mathrm{tr}\!\left[
\gamma_{\mu}\gamma^{i}\gamma_{\nu}\gamma^{j}P_{R}
\right]
\hat{P}_{s}^{\mu}\hat{R}_{t}^{\nu}.
\end{equation}
At this point the Dirac algebra can be reduced completely. Using \(P_{R}=(1+\gamma_{5})/2\), \(\epsilon^{0123}=+1\), and the same metric convention as in the main text, the relevant identities are
\begin{equation}
\epsilon_{ij3}\,
\mathrm{tr}\!\left[
\gamma_{\mu}\gamma^{i}\gamma_{\nu}\gamma^{0}
\right]
a^{\mu}b^{\nu}
=
4\,\epsilon_{ij3}\left(a^{i}b^{0}+a^{0}b^{i}\right),
\end{equation}
\begin{equation}
\epsilon_{ij3}\,
\mathrm{tr}\!\left[
\gamma_{\mu}\gamma^{i}\gamma_{\nu}\gamma^{0}\gamma_{5}
\right]
a^{\mu}b^{\nu}
=
4i\left(a^{j}b^{3}-a^{3}b^{j}\right),
\end{equation}
\begin{equation}
\epsilon_{ij3}\,
\mathrm{tr}\!\left[
\gamma_{\mu}\gamma^{i}\gamma_{\nu}\gamma^{j}
\right]
a^{\mu}b^{\nu}
=
0,
\end{equation}
\begin{equation}
\epsilon_{ij3}\,
\mathrm{tr}\!\left[
\gamma_{\mu}\gamma^{i}\gamma_{\nu}\gamma^{j}\gamma_{5}
\right]
a^{\mu}b^{\nu}
=
8i\left(a^{3}b^{0}-a^{0}b^{3}\right).
\end{equation}
Consequently,
\begin{equation}
\epsilon_{ij3}\,
\mathrm{tr}\!\left[
\gamma_{\mu}\gamma^{i}\gamma_{\nu}\gamma^{0}P_{R}
\right]
a^{\mu}b^{\nu}
=
2\,\epsilon_{ij3}\left(a^{i}b^{0}+a^{0}b^{i}\right)
+
2i\left(a^{j}b^{3}-a^{3}b^{j}\right),
\end{equation}
while
\begin{equation}
\epsilon_{ij3}\,
\mathrm{tr}\!\left[
\gamma_{\mu}\gamma^{i}\gamma_{\nu}\gamma^{j}P_{R}
\right]
a^{\mu}b^{\nu}
=
4i\left(a^{3}b^{0}-a^{0}b^{3}\right).
\end{equation}
These relations show explicitly that the \((0j)\) piece contains both the ordinary four-gamma contraction and the chiral \(\gamma_{5}\) contraction before the angular integral is carried out, whereas the \((j0)\) piece is purely parity-odd after contraction with \(\epsilon_{ij3}\). This is precisely why the decomposition is useful: the part weighted by \(P^{j}\) and the part weighted by \(i\tilde{\omega}_{n}\) have different tensor structure before the static limit, even though both ultimately contribute to the same parity-odd transport coefficient. At the same time, the GUP factor \(W_{\beta}(P)\) multiplies both contributions as a function of \(P^{2}\) alone, so it changes the radial momentum weighting but does not alter the decomposition itself. 
{The source convention is the one specified in Sec.~\ref{sec3}. 
Possible source-dependent contributions beyond the DoS
sector are discussed in Sec.~\ref{sec:consistency_anomalies}.
}
Equations above are therefore the complete decomposition formulae needed for the next stage of the calculation, where the parity-odd projection, Matsubara sum, vacuum subtraction, and static long-wavelength limit reduce the correlator to the one-dimensional representation used in the main text.

\subsection{Parity-odd projection and rotational structure}
\label{A.2}
The purpose of this subsection is to make explicit the tensor reduction that lies behind the Kubo formula used in the main paper. 
In the present one-loop analysis, \(W_\beta(P)\) appears solely as a
rotationally invariant scalar multiplying the spatial momentum
integral, while the current and stress-tensor vertices retain their
undeformed massless-chiral-theory forms. The parity-odd sector of the
mixed Euclidean correlator is consequently determined by the chiral
Dirac trace and the angular structure of the loop integral. 
The key point is that, in a rotationally invariant equilibrium state, the antisymmetric part of a two-index spatial correlator at first order in the external momentum can only be proportional to $i\varepsilon^{ij\ell}k_\ell$, with $\varepsilon^{abc}\equiv \epsilon^{0abc}$ the three-dimensional Levi-Civita tensor. The derivation below shows this explicitly and also demonstrates why the GUP factor $W_\beta(P)$ can modify only the scalar coefficient multiplying that structure, not the structure itself \cite{Landsteiner:2011cp,Landsteiner:2012kd,Son:2009tf}. We write the Euclidean external momentum and loop momentum as
\begin{equation}
K=(i\omega_m,\mathbf{k}),\qquad P=(i\tilde\omega_n,\mathbf{P}),\qquad P\equiv |\mathbf{P}|,
\end{equation}
and we denote
\begin{equation}
D_n(P)\equiv \tilde\omega_n^2+P^2,\qquad
D_{n+m}(P+K)\equiv (\tilde\omega_n+\omega_m)^2+(\mathbf{P}+\mathbf{k})^2.
\end{equation}
The mixed correlator may then be written as
\begin{equation}
G_E^{ij}(K)
=
T\sum_{n\in\mathbb Z}\int_{\mathbf P} W_\beta(P)\,
\frac{\mathcal N^{ij}(P,K)}{D_n(P)\,D_{n+m}(P+K)},
\qquad
\int_{\mathbf P}\equiv \int \frac{d^3P}{(2\pi)^3},
\end{equation}
with
\begin{equation}
\mathcal N^{ij}(P,K)=\mathcal N^{ij}_{(0j)}(P,K)+\mathcal N^{ij}_{(j0)}(P,K),
\end{equation}
where the two pieces correspond to the $\gamma^0(2P+K)^j/2$ and $\gamma^j(2i\tilde\omega_n+i\omega_m)/2$ parts of the $T^{0j}$ vertex. 
{Using the operator definitions and Dirac-trace identities derived in
App.~\ref{A.1}, the parity-odd numerators reduce to}
\begin{align}
\mathcal N^{ij}_{(0j),\mathrm{odd}}(P,K)
&=
(2P+K)^j\,\epsilon^{i\alpha 0\beta}\,P_\alpha (P+K)_\beta,
\label{A3N0j}\\[4pt]
\mathcal N^{ij}_{(j0),\mathrm{odd}}(P,K)
&=
(2i\tilde\omega_n+i\omega_m)\,\epsilon^{i\alpha j\beta}\,P_\alpha (P+K)_\beta.
\label{A3Nj0}
\end{align}
At this stage the first crucial simplification appears. Since $\epsilon^{i\alpha\mu\beta}$ is antisymmetric in $(\alpha,\beta)$,
\begin{equation}
\epsilon^{i\alpha\mu\beta}P_\alpha P_\beta=0,
\end{equation}
and therefore
\begin{equation}
\epsilon^{i\alpha\mu\beta}P_\alpha(P+K)_\beta
=
\epsilon^{i\alpha\mu\beta}P_\alpha K_\beta.
\label{A3linearK}
\end{equation}
Thus every parity-odd term is automatically linear in the external momentum. In particular, the parity-odd numerator vanishes identically at $K=0$, which is exactly what one expects on general grounds: without an external spatial direction there is no pseudovector available from which to build an antisymmetric spatial two-tensor.

To isolate the leading vortical structure, we now take the static limit $\omega_m\to 0$ and retain only terms linear in $\mathbf{k}$. Since the numerator is already $O(k)$ by Eq.~\eqref{A3linearK}, the denominator may be replaced by its zeroth-order isotropic form,
\begin{equation}
\frac{1}{D_n(P)\,D_{n+m}(P+K)}
=
\frac{1}{D_n(P)^2}+O(k,\omega_m),
\end{equation}
because any $O(k)$ correction from the denominator would contribute only at $O(k^2)$ to the parity-odd correlator. In the static limit, Eq.~\eqref{A3N0j} becomes
\begin{equation}
\mathcal N^{ij}_{(0j),\mathrm{odd}}(P,K)\Big|_{\omega_m=0}
=
(2P+K)^j\,\varepsilon^{i a b}\,P_a k_b,
\label{A3firststatic}
\end{equation}
where $a,b\in\{1,2,3\}$ are spatial indices. The term proportional to $K^j\varepsilon^{iab}P_a k_b$ is odd under $\mathbf P\to -\mathbf P$ and vanishes after angular integration, so only the $2P^j$ part survives. For any angular-independent function $F(P)$ one has the standard isotropic identities
\begin{equation}
\int d\Omega\, P^a\,F(P)=0,
\quad
\int d\Omega\, P^a P^b\,F(P)=\frac{4\pi}{3}P^2\delta^{ab}F(P),
\quad
\int d\Omega\, P^a P^b P^c\,F(P)=0.
\label{A3isotropic}
\end{equation}
Using Eq.~\eqref{A3isotropic}, the angular average of Eq.~\eqref{A3firststatic} is
\begin{align}
\int d\Omega\,\mathcal N^{ij}_{(0j),\mathrm{odd}}(P,K)\Big|_{\omega_m=0}
&=
2\int d\Omega\,P^j\varepsilon^{iab}P_a k_b
\nonumber\\[2pt]
&=
2\,\frac{4\pi}{3}P^2\,\delta^{ja}\varepsilon^{iab}k_b
\nonumber\\[2pt]
&=
\frac{8\pi}{3}P^2\,\varepsilon^{ijb}k_b.
\label{A3angfirst}
\end{align}
This already shows explicitly how the first stress-tensor piece collapses onto the unique pseudotensor $\,\varepsilon^{ijb}k_b$. 
The second piece, Eq.~\eqref{A3Nj0}, is slightly different because the factor multiplying the trace is a scalar in spatial indices. Put $\omega_m=0$ and decomposing the index $\alpha$ into time and space components gives
\begin{align}
\mathcal N^{ij}_{(j0),\mathrm{odd}}(P,K)\Big|_{\omega_m=0}
&=
2i\tilde\omega_n\,\epsilon^{i0jb}\,(i\tilde\omega_n)\,k_b
+
2i\tilde\omega_n\,\epsilon^{iajb}\,P_a k_b
\nonumber\\[2pt]
&=
2\tilde\omega_n^2\,\varepsilon^{ijb}k_b
+
2i\tilde\omega_n\,\epsilon^{iajb}P_a k_b .
\label{A3secondstatic}
\end{align}
The second term in Eq.~\eqref{A3secondstatic} is again odd in $\mathbf P$ and vanishes after angular integration. The first term already has the desired tensor form and is angle-independent. Therefore
\begin{equation}
\int d\Omega\,\mathcal N^{ij}_{(j0),\mathrm{odd}}(P,K)\Big|_{\omega_m=0}
=
8\pi\,\tilde\omega_n^2\,\varepsilon^{ijb}k_b.
\label{A3angsecond}
\end{equation}
Combining Eqs.~\eqref{A3angfirst} and \eqref{A3angsecond}, the parity-odd part of the full mixed correlator takes the form
\begin{equation}
G^{ij}_{E,\mathrm{odd}}(K)
=
i\,\varepsilon^{ijb}k_b\,
T\sum_{n\in\mathbb Z}
\int_0^\infty \frac{dP\,P^2}{2\pi^2}\,
W_\beta(P)\,
\mathcal F(P,\tilde\omega_n;\omega_m,k^2),
\label{A3generalform}
\end{equation}
for some scalar kernel $\mathcal F$ determined by the remaining radial and Matsubara algebra. The precise closed form of $\mathcal F$ is not needed in the present subsection; what matters here is that all dependence on the loop direction $\hat{\mathbf P}$ has disappeared, and the entire parity-odd sector has factorized into the unique rotationally allowed pseudotensor $i\varepsilon^{ijb}k_b$ multiplying a scalar function of $(\omega_m,k^2,\mu,T,\beta)$.

Eq.~\eqref{A3generalform} can also be justified directly from symmetry. In an isotropic equilibrium state the only available building blocks are the spatial Kronecker delta $\delta^{ij}$, the Levi-Civita tensor $\varepsilon^{ij\ell}$, the external momentum $k^\ell$, and scalar functions of $k^2$, $\omega_m$, $T$, and $\mu$. An antisymmetric rank-two spatial tensor linear in $\mathbf{k}$ can therefore only be
\begin{equation}
G^{ij}_{E,\mathrm{odd}}(K)
=
i\,\varepsilon^{ij\ell}k_\ell\,\Pi_A(i\omega_m,k^2,\mu,T,\beta),
\label{A3decomp}
\end{equation}
with a scalar coefficient $\Pi_A$. Terms proportional to $i\varepsilon^{ij\ell}P_\ell$ are forbidden after loop integration because they are odd in $\mathbf P$, while terms with more powers of $\mathbf P$ reduce, by isotropy, to $\delta$-tensors contracted into $\varepsilon^{ij\ell}k_\ell$ exactly as in Eq.~\eqref{A3angfirst}. No other independent pseudotensor exists at this order. 
{The factor \(W_\beta(P)\), defined in
Eq.~\eqref{eq:LiouvilleMeasure}, depends only on \(P^2\) and
therefore commutes with the angular reduction.
}
For any sufficiently regular scalar function $F(P)$,
\begin{align}
\int_{\mathbf P} W_\beta(P)\,P^a\,F(P)&=0,\\[2pt]
\int_{\mathbf P} W_\beta(P)\,P^aP^b\,F(P)
&=
\frac{\delta^{ab}}{3}
\int_{\mathbf P} W_\beta(P)\,P^2\,F(P),\\[2pt]
\int_{\mathbf P} W_\beta(P)\,P^aP^bP^c\,F(P)&=0,
\end{align}
and similarly for higher angular moments.  Its dependence
is therefore confined to the scalar coefficient
\(\Pi_A\) multiplying the unique antisymmetric structure
\(i\epsilon^{ij\ell}k_\ell\). The tensor basis of the parity-odd
correlator is consequently identical to that of the undeformed Weyl
theory.  
Choosing the external momentum along the \(z\)-axis,
\(\bm{k}=k\,\hat{\bm z}\), Eq.~\eqref{A3decomp}
reduces to
\begin{equation}
G^{ij}_{E,\mathrm{odd}}(0,k\hat z)
=
i\,\varepsilon^{ij3}\,k\,\Pi_A(0,0,\mu,T,\beta)+O(k^2),
\end{equation}
which is the form used in the main text and in Appendix \ref{A.3}. The determination of the chiral vortical conductivity therefore reduces to the scalar coefficient \(\Pi_A\). The parity-odd tensor projection and its insensitivity to the isotropic GUP weight have already been established exactly at the level of the angular reduction.

\subsection{Matsubara sum, vacuum subtraction, and static limit}
\label{A.3}
{ We now complete the one-loop evaluation of the mixed
current-momentum density correlator by carrying out the Matsubara
sums, removing the temperature- and chemical-potential independent
vacuum contribution, and taking the ordered static,
long-wavelength limit explicitly.} Throughout this subsection we keep the conventions of Appendix~\ref{A.1} and
Appendix~\ref{A.2}, choose the external spatial momentum along the rotation axis,
\begin{equation}
\mathbf{k}=k\,\hat z,
\end{equation}
denote
\begin{equation}
P\equiv |\mathbf{P}|,\qquad P'\equiv |\mathbf{P}+\mathbf{k}|=\sqrt{P^2+k^2+2Pkx},\qquad x\equiv \cos\theta,
\end{equation}
{and use the DoS factor defined in Eq.~\eqref{eq:LiouvilleMeasure}. Since \(W_\beta(P)\) carries no
Matsubara-frequency dependence, it leaves the contour structure of
the thermal sum unchanged. The source convention is the same as in
Sec.~\ref{sec3}, with its source-covariant extension discussed in
Sec.~\ref{sec:consistency_anomalies}.}
The traced expressions obtained in Appendix~\ref{A.2} reduce the frequency dependence to two basic Matsubara kernels. For the piece proportional to \(\gamma^0 P^j\) in the stress-tensor vertex we define
{\begin{equation}
S^{tu}_0(i\omega_m,P,P')
\equiv
T\sum_{n\in\mathbb{Z}}
\Delta_t(i\widetilde{\omega}_n,P)
\Delta_u(i\widetilde{\omega}_n+i\omega_m,P'),
\qquad
\Delta_t(z,P)\equiv\frac{1}{z-tP},
\qquad
t,u=\pm,
\end{equation}
while for the piece proportional to \(\gamma^j i\tilde\omega_n\) we define
\begin{equation}
S^{tu}_1(i\omega_m,P,P')
\equiv
T\sum_{n\in\mathbb{Z}}
i\widetilde{\omega}_n\,
\Delta_t(i\widetilde{\omega}_n,P)
\Delta_u(i\widetilde{\omega}_n+i\omega_m,P').
\label{eq:S1kernel}
\end{equation}
The shifted fermionic frequencies are
$i\widetilde{\omega}_n=i(2n+1)\pi T+\mu$, while
$\omega_m=2\pi mT$ is the bosonic Matsubara frequency, so that the
corresponding Euclidean external energy is $i\omega_m$. The Matsubara
sums are evaluated using the standard contour formula
\begin{equation}
T\sum_{n\in\mathbb{Z}}F(i\widetilde{\omega}_n)
=
-\sum_{z_\star\in{\rm poles}(F)}
n_F(z_\star-\mu)\,
{\rm Res}_{z=z_\star}F(z),
\qquad
n_F(y)\equiv\frac{1}{e^{y/T}+1}.
\label{eq:MatsubaraContour}
\end{equation}}
Applying this to \(\mathcal{S}_0^{tu}\), whose poles are at \(z=tP\) and \(z=uP'-i\omega_m\), and using the bosonic periodicity
\begin{equation}
n_F(uP'-i\omega_m-\mu)=n_F(uP'-\mu),
\end{equation}
one obtains
\begin{equation}
\mathcal{S}_0^{tu}(i\omega_m,P,P')
=
\frac{n_F(tP-\mu)-n_F(uP'-\mu)}{i\omega_m+tP-uP'}.
\end{equation}
To express this in terms of positive-energy Fermi functions, we use
\begin{equation}
n_F(tP-\mu)=\frac{1-t}{2}+t\,n_F(P-t\mu),
\qquad
n_F(uP'-\mu)=\frac{1-u}{2}+u\,n_F(P'-u\mu),
\end{equation}
which gives
\begin{equation}
\mathcal{S}_0^{tu}(i\omega_m,P,P')
=
\frac{
t\,n_F(P-t\mu)-u\,n_F(P'-u\mu)+\frac12(u-t)
}{
i\omega_m+tP-uP'
}.
\label{eq:A4-S0-full}
\end{equation}
The last term in the numerator,
\begin{equation}
\frac{\frac12(u-t)}{i\omega_m+tP-uP'},
\end{equation}
is independent of \(T\) and \(\mu\), and therefore belongs to the vacuum part of the correlator. 
For \(\mathcal{S}_1^{tu}\) the same contour method yields
\begin{equation}
\mathcal{S}_1^{tu}(i\omega_m,P,P')
=
\frac{
tP\,n_F(tP-\mu)-(uP'-i\omega_m)\,n_F(uP'-\mu)
}{
i\omega_m+tP-uP'
}.
\end{equation}
Using again
\begin{equation}
tP\,n_F(tP-\mu)=P\,n_F(P-t\mu)-\frac{1-t}{2}P,
\end{equation}
and
\begin{equation}
(uP'-i\omega_m)\,n_F(uP'-\mu)
=
(P'-u\,i\omega_m)\,n_F(P'-u\mu)
+\frac{1-u}{2}(uP'-i\omega_m),
\end{equation}
we separate the thermal and vacuum pieces as
\begin{equation}
\mathcal{S}_1^{tu}(i\omega_m,P,P')
=
\frac{
P\,n_F(P-t\mu)-(P'-u\,i\omega_m)\,n_F(P'-u\mu)
}{
i\omega_m+tP-uP'
}
-
\frac{
\frac12(1-t)P+\frac12(1-u)(uP'-i\omega_m)
}{
i\omega_m+tP-uP'
}.
\label{eq:A4-S1-full}
\end{equation}
The second term is again independent of \(T\) and \(\mu\), and is therefore part of the vacuum contribution. 
It is convenient to define the vacuum-subtracted odd correlator by
\begin{equation}
\widehat G^{ij}_{E,\mathrm{odd}}(i\omega_m,\mathbf{k})
\equiv
G^{ij}_{E,\mathrm{odd}}(i\omega_m,\mathbf{k})
-
G^{ij}_{E,\mathrm{odd}}(i\omega_m,\mathbf{k})\big|_{T=\mu=0}.
\end{equation}
At the level of the Matsubara kernels this means that only the thermal pieces of \eqref{eq:A4-S0-full} and \eqref{eq:A4-S1-full} are retained:
\begin{equation}
\overline{\mathcal{S}}_0^{tu}(i\omega_m,P,P')
=
\frac{
t\,n_F(P-t\mu)-u\,n_F(P'-u\mu)
}{
i\omega_m+tP-uP'
},
\label{eq:A4-S0-thermal}
\end{equation}
\begin{equation}
\overline{\mathcal{S}}_1^{tu}(i\omega_m,P,P')
=
\frac{
P\,n_F(P-t\mu)-(P'-u\,i\omega_m)\,n_F(P'-u\mu)
}{
i\omega_m+tP-uP'
}.
\label{eq:A4-S1-thermal}
\end{equation}
Because \(n_F(P\mp\mu)\) decays exponentially at large \(P\), the vacuum-subtracted thermal integrals are ultraviolet finite. The only ultraviolet-divergent piece is the zero-temperature, zero-density vacuum term that has been removed explicitly, exactly as in the standard free-fermion Kubo derivation \cite{Landsteiner:2012kd,Landsteiner:2011cp}. 

Substituting Eqs.~\eqref{eq:A4-S0-thermal} and \eqref{eq:A4-S1-thermal} into the traced parity-odd expressions
from Appendix~\ref{A.2}, one may perform the \(t,u=\pm\) sums before adding the
two Belinfante-vertex contributions. This gives a useful intermediate check
on the normalization. The term proportional to \(\gamma^0 P^j\) gives, in the
static long-wavelength limit,
\begin{equation}
\widehat G^{ij}_{E,\mathrm{odd},(0j)}(0,k\hat z)
=
i\epsilon^{ij3}k\,
\frac{1}{4\pi^2}
\int_0^\infty dP\,P\,W_\beta(P)
\left[n_F(P-\mu)+n_F(P+\mu)\right]
+O(k^2),
\label{eq.173}
\end{equation}
where the hat denotes the vacuum-subtracted correlator. The term proportional
to \(\gamma^j i\widetilde\omega_n\) gives the same static contribution,
\begin{equation}
\widehat G^{ij}_{E,\mathrm{odd},(j0)}(0,k\hat z)
=
i\epsilon^{ij3}k\,
\frac{1}{4\pi^2}
\int_0^\infty dP\,P\,W_\beta(P)
\left[n_F(P-\mu)+n_F(P+\mu)\right]
+O(k^2).
\label{eq.174}
\end{equation}
The equality of Eqs.~\eqref{eq.173} and \eqref{eq.174} follows directly from the on-shell identity
\begin{equation}
i\widetilde\omega_n\,\Delta_t(i\widetilde\omega_n,P)
=
1+tP\,\Delta_t(i\widetilde\omega_n,P).
\label{eq.175}
\end{equation}
After the vacuum subtraction, the first term on the right-hand side of
Eq.~\eqref{eq.175} contributes only to the removed temperature- and chemical potential 
independent contact part. The remaining on-shell term \(tP\Delta_t\) produces
the same angular kernel and the same symmetrized thermal combination
\begin{equation}
f_V(P)\equiv n_F(P-\mu)+n_F(P+\mu)
\label{eq.176}
\end{equation}
as the \(\gamma^0P^j\) part of the Belinfante vertex. 
For completeness, we spell out the ordered static limit used in obtaining
Eqs.~\eqref{eq.173} and \eqref{eq.174}. After the Matsubara sum and analytic continuation, the angular dependence is carried by \(P'=|\mathbf P+\mathbf k|,\) with
\begin{equation}
P'=\sqrt{P^2+k^2+2Pkx}=P+kx+O(k^2),
\quad x=\cos\theta .
\label{eq.177}
\end{equation}
The parity-odd angular kernel reduces to
\begin{equation}
\frac{P^2 k-(\mathbf P\cdot \mathbf k)P_z}{PP'}
=
k(1-x^2)+O(k^2),
\label{eq.178}
\end{equation}
and hence
\(
\int d\Omega\,(1-x^2)=\frac{8\pi}{3}.
\)
Equivalently, the same limit is encoded in the standard weak-coupling identity
\begin{equation}
\lim_{k\to0}\lim_{\omega\to0}
\sum_{t=\pm}
\log\left[
\frac{\Omega_t^2-(P+k)^2}{\Omega_t^2-(P-k)^2}
\right]
=
\frac{2k}{P},
\qquad
\Omega_t=\omega+i0^+ +tP .
\label{eq.179}
\end{equation}
The GUP factor does not modify this limiting operation. It depends only on
the radial variable \(P\), carries no Matsubara-frequency dependence, and is
independent of the angular variable \(x\) and of the helicity labels \(t,u\).
Therefore it commutes with the \(t,u\) sums, the angular integration, and the
ordered static limit. After vacuum subtraction the thermal factor
\(f_V(P)\) decays exponentially at large \(P\), so multiplication by
\(W_\beta(P)\) leaves the radial integral convergent and changes only the
final radial weight. 
{The remaining radial integral is precisely \(I_\beta(\mu,T)\) defined in Eq.~\eqref{eq:Ibeta}.} 
Eqs.~\eqref{eq.173} and \eqref{eq.174} become
\begin{equation}
\widehat G^{ij}_{E,\mathrm{odd},(0j)}(0,k\hat z)
=
\frac{i}{2}\epsilon^{ij3}k\,I_\beta(\mu,T)+O(k^2),
\end{equation}
and
\begin{equation}
\widehat G^{ij}_{E,\mathrm{odd},(j0)}(0,k\hat z)
=
\frac{i}{2}\epsilon^{ij3}k\,I_\beta(\mu,T)+O(k^2).
\end{equation}
{ The two contributions are therefore equal, and their sum gives the
vacuum-subtracted parity-odd correlator
\begin{equation}
\widehat{G}^{ij}_{E,\mathrm{odd}}
(0,k\hat{\bm z})
=
i\epsilon^{ij3}k\,I_\beta(\mu,T)
+\mathcal O(k^2).
\label{eq:appendix_Godd_final}
\end{equation}
}
Analytic continuation relates the retarded and Euclidean correlators by
\begin{equation}
G_R^{ij}(\omega,\mathbf{k})
=
-\,G_E^{ij}(i\omega_m\to \omega+i0^+,\mathbf{k}),
\end{equation}
Substituting Eq.~\eqref{eq:appendix_Godd_final} into the Kubo relation gives
\begin{align}
\sigma_V^{\mathrm{GUP}}(\mu,T,\beta)
&=
\lim_{k\to0}\frac{i}{2k}\epsilon_{ij3}\,
G_R^{ij}(0,k\hat z)
\nonumber\\
&=
\lim_{k\to0}\frac{i}{2k}\epsilon_{ij3}
\Bigl[-\,\overline{G}^{\,ij}_{E,\mathrm{odd}}(0,k\hat z)\Bigr]
\nonumber\\
&=
\lim_{k\to0}\frac{i}{2k}\epsilon_{ij3}
\Bigl[-\,i\,\epsilon^{ij3}k\,I_\beta(\mu,T)\Bigr]
\nonumber\\
&=
I_\beta(\mu,T),
\end{align}
{
here \(\epsilon^{ij3}\epsilon^{ij3}=2\). This reproduces the one-dimensional conductivity in Eq.~\eqref{eq:sigmaV_integral_GUP}. The Matsubara sum yields the symmetric thermal combination \(n_F(P-\mu)+n_F(P+\mu)\), while the temperature- and chemical-potential-independent vacuum contribution is removed before the radial integration. The undeformed  limit and the corresponding anomaly interpretation are discussed in Secs.~\ref{sec3} and
\ref{sec:consistency_anomalies}.
}

\end{document}